\documentclass[a4paper,11pt,preprintnumbers]{article}
\pdfoutput=1

\usepackage{amssymb,mathrsfs,amsmath}
\usepackage{a4wide}
\usepackage{color}
\usepackage[dvipsnames,table]{xcolor}
\usepackage{slashed,soul}
\usepackage{graphicx}
\usepackage{amsfonts}
\usepackage{lscape}
\def\linkcolor{cyan!70!black}

\usepackage[
colorlinks=true
,urlcolor=\linkcolor
,anchorcolor=\linkcolor
,citecolor=\linkcolor
,filecolor=\linkcolor
,linkcolor=\linkcolor
,menucolor=\linkcolor
,linktocpage=true
,pdfproducer=medialab
,pdfa=true
]{hyperref}
\usepackage{amsthm}
\usepackage{booktabs}
\usepackage{array}
\usepackage{rotating}
\usepackage[numbers, sort&compress]{natbib}
\usepackage{multirow}
\usepackage{float}
\usepackage[utf8]{inputenc}
\usepackage[T1]{fontenc}
\usepackage{appendix}
\usepackage{epsfig}
\usepackage{orcidlink}
\usepackage{comment}
\usepackage{adjustbox}
\usepackage{tcolorbox}
\usepackage{booktabs}

\newcolumntype{C}[1]{>{\centering\arraybackslash$} wc{#1} <{$}}

\newcommand{\beq}{\begin{equation}} 
\newcommand{\eeq}{\end{equation}} 
\newcommand{\ba}{\begin{array}}  
\newcommand{\ea}{\end{array}} 
\newcommand{\bea}{\begin{eqnarray}}  
\newcommand{\eea}{\end{eqnarray} }  
\newcommand{\bal}{\begin{align}}
\newcommand{\eal}{\end{align}}   
\newcommand{\bi}{\begin{itemize}}  
\newcommand{\ei}{\end{itemize}}  
\newcommand{\ben}{\begin{enumerate}}  
\newcommand{\een}{\end{enumerate}}  
\newcommand{\bc}{\begin{center}}
\newcommand{\ec}{\end{center}} 
\newcommand{\bt}{\begin{table}}
\newcommand{\et}{\end{table}}  
\newcommand{\btb}{\begin{tabular}}
\newcommand{\etb}{\end{tabular}}

\newcommand{\hc}{\mathrm{h.c.}}
\renewcommand{\O}{\mathcal{O}}
\newcommand{\op}[3]{\O^{#2,#3}_{#1}}
\newcommand{\opd}[2]{\O^{#2}_{#1}}
\newcommand{\wcL}[2]{c^{#2}_{#1}} 
\newcommand{\wcH}[2]{C^{#2}_{#1}} 

\newcommand{\abs}[1]{\left\lvert#1\right\rvert}

\def\arrvline{\hfil\kern\arraycolsep\vline\kern-\arraycolsep\hfilneg}
\newcommand{\sw}{s_{\mathrm{w}}}
\newcommand{\cw}{c_{\mathrm{w}}}

\newcommand{\calO}{{\mathcal{O}}}

\newcommand{\hj }{\mbox{${H^\dag\, \raisebox{1.5mm}{${}^\leftrightarrow$}\hspace{-4mm} D_\mu\,H}$}}

\allowdisplaybreaks

\let\OLDthebibliography\thebibliography
\renewcommand\thebibliography[1]{
  \OLDthebibliography{#1}
  \setlength{\parskip}{0pt}
  \setlength{\itemsep}{0pt plus 0.3ex}
}

\begin{document}

\vspace{1cm}

\begin{titlepage}

\begin{flushright}
IFT-UAM/CSIC-24-39
 \end{flushright}
\vspace{0.2truecm}

\begin{center}
\renewcommand{\baselinestretch}{1.8}\normalsize
\boldmath
{\LARGE\textbf{
Global Lepton Flavour Violating Constraints \\ on New Physics
}}
\unboldmath
\end{center}

\vspace{0.4truecm}

\renewcommand*{\thefootnote}{\fnsymbol{footnote}}

\begin{center}

{
Enrique Fern\'andez-Mart\'inez\footnote{\href{mailto:enrique.fernandez@csic.es}{enrique.fernandez@csic.es}}\orcidlink{0000-0002-6274-4473}, 
Xabier Marcano\footnote{\href{mailto:xabier.marcano@uam.es}{xabier.marcano@uam.es}}\orcidlink{0000-0003-0033-0504}
and Daniel Naredo-Tuero\footnote{\href{mailto:daniel.naredo@uam.es}{daniel.naredo@uam.es}}\orcidlink{0000-0002-5161-5895}
}

\vspace{0.7truecm}

{\footnotesize
Departamento de F\'{\i}sica Te\'orica and Instituto de F\'{\i}sica Te\'orica UAM/CSIC,\\
Universidad Aut\'onoma de Madrid, Cantoblanco, 28049 Madrid, Spain
}

\vspace*{2mm}
\end{center}

\renewcommand*{\thefootnote}{\arabic{footnote}}
\setcounter{footnote}{0}

\begin{abstract}

We perform a global analysis of the bounds from charged lepton flavour violating observables to new physics. We parametrize generic new physics through the Effective Field Theory formalism and perform global fits beyond the common one-operator-at-a-time analyses to investigate how much present data is able to constrain the full parameter space. We particularly focus on leptonic and semileptonic operators with light quarks, identifying unbounded flat directions, detailing how many are present and which operators are involved. The analysis is performed in the general LEFT formalism, which contains all possible low-energy effective operators relevant for lepton flavour violation, as well as in more restricted scenarios, when operators come from a SMEFT completion. We find that flat directions play no role in the fully leptonic four-fermion operators. Conversely, they significantly hinder the ability to derive global bounds on semileptonic operators, with several flat or at least very poorly constrained directions preventing to fully constrain the parameter space. These results are particularly affected by the proper inclusion of uncertainties in the parameters describing $\mu-e$ conversion, which decrease the number of well-constrained directions in operator space when treated as nuisance parameters in the fit. While present data is able to provide global constraints on all operators only in the more restricted scenarios we investigated, very strong correlations among the parameters must exist to avoid conflict with the different observables. We provide correlation matrices approximating our full results as a useful tool to compare present data with particular UV completions.

\end{abstract}

\end{titlepage}

\tableofcontents

\section{Introduction}

The Effective Field Theory (EFT) formalism constitutes an extremely useful tool to parametrize new physics beyond the Standard Model (SM) of particle physics and to derive model-independent constrains on its existence. In particular, when the new particles and interactions of the theory beyond the SM are characterized by a mass scale that is not yet achievable in our present searches, it is very natural to integrate these new heavy degrees of freedom out of the theory. Then, their indirect effects at low energies are instead represented by a tower of effective operators of dimension larger than 4 involving the light degrees of freedom and suppressed by corresponding powers of the heavy mass scale through their Wilson coefficients.      

Among all the observables testing the validity of the SM and probing for new physics, flavour changing processes are extremely suppressed in the SM due to the Glashow-Iliopoulus-Maiani (GIM) mechanism~\cite{Glashow:1970gm} and, as such, provide one of our best windows to explore the physics beyond. This fact is particularly true for charged lepton flavour violating (cLFV) processes, where the GIM cancellation is controlled by the negligibly small neutrino masses. Taken at face value, processes such as $\mu \to e \gamma$ or $\mu-e$ conversion in nuclei are constraining the mass scale characterizing flavour-changing $d=6$ effective operators to $\Lambda > 10^3$~TeV~\cite{Calibbi:2017uvl,Davidson:2022jai}, far beyond the reach of collider searches. 

The SMEFT~\cite{Buchmuller:1985jz,Grzadkowski:2010es}, that is the EFT that can be built with the SM particle content and respecting its symmetries, is characterized by a very large number of operators, particularly when a general flavour structure allowing for flavour violation is allowed. In this scenario, 2499 parameters are necessary to describe the SMEFT at $d=6$~\cite{Alonso:2013hga}, the lowest dimension inducing cLFV transitions. Given the complexity of the problem, one of two common simplifying assumptions is usually made when studying the constraints existing in the EFT operators. The first is to keep only terms linear in the Wilson coefficients of the effective operators, that is, keeping only the interference with the SM contributions~\cite{Falkowski:2017pss,Falkowski:2020pma,Brivio:2022hrb}. This approach allows to perform comprehensive global fits and identify possible flat directions that may avoid constraints. On the other hand, this approach is not appropriate when dealing with processes that are not possible or very suppressed in the SM, such as charged lepton flavour changing processes. In these cases a common simplification is to consider only one operator at a time~\cite{Raidal:1997hq,Kuno:1999jp,Brignole:2004ah,Cirigliano:2009bz,Crivellin:2013hpa,Celis:2014asa,Crivellin:2017rmk,Cirigliano:2017azj,Davidson:2017nrp,Falkowski:2021bkq,Cirigliano:2021img,Ardu:2021koz,Calibbi:2021pyh,Kumar:2021yod,Calibbi:2022ddo,Plakias:2023esq,Ardu:2023yyw,Ardu:2024bua}.
While this allows for a very straightforward derivation of constraints for the corresponding Wilson coefficients, these bounds may be regarded as too aggressive, since they will miss any possible flat directions between the different effective operators contributing to a given observable. 

In this work we explore this issue beyond the one-operator-at-a-time approach aiming to investigate what is the present status of the constraints that can be placed on effective operators violating charged lepton flavour. We perform a global analysis of both low-energy EFT (LEFT) and SMEFT frameworks, extending previous results for the LEFT $\mu-e$ sector~\cite{Davidson:2022nnl,Hoferichter:2022mna} and SMEFT $\tau-\ell$ sector~\cite{Husek:2020fru,Banerjee:2022xuw}. 
As a simplifying first step, for the operators that contribute to cLFV with an hadronic component, we do not consider flavour change in the quark sector and we focus only on interactions with the lighter quarks ($u$, $d$ and $s$), relevant for the most important observables. Moreover, we will derive constraints on the Wilson coefficients of effective operators directly at the low-energy scale relevant for the observables. 

In section~\ref{sec:EFT} we first list the LEFT operators that may contribute to the cLFV observables and discuss how the matching with the SMEFT affects how many independent operators may appear in total. In section~\ref{sec:lepobs} we describe the fully leptonic cLFV observables and conclude that, in this case, the one-at-a-time approach would be equivalent to a more elaborate global fit involving several operators. In section~\ref{sec:obs} we describe the relevant observables constraining semileptonic operators and detail how many independent combinations can be probed, highlighting the importance of nuclear uncertainties when exploring the $\mu-e$ sector. In section~\ref{sec:LEFT} we discuss if any semileptonic LEFT operators are already constrained by present data when cancellations among different operators are allowed and  discuss the existence of flat directions. In section~\ref{sec:SMEFT} we study how these results are affected by the matching to the SMEFT. In section~\ref{sec:SMEFT-s} we make the final simplifying assumption of considering only first generation quarks in the interactions and discuss the conditions under which bounds for all operators may be derived. We summarize our results and present our conclusions in section~\ref{sec:concl}.
Finally, we provide the details of our analysis as well as the correlation matrices in the appendices, so our global constrains can be easily incorporated to particular UV-complete scenarios.

\section{EFT framework}
\label{sec:EFT}

\begin{table}[t!] 
\centering
\renewcommand{\arraystretch}{1.2}
\setlength{\tabcolsep}{15pt}
\begin{tabular}{lclc} 
\hline
\multicolumn{4}{c}{$2q2\ell$ operators} \\
\hline \\[-2.6ex]
$\calO_{\ell q,\alpha \beta \gamma \delta}^{(1)}$ & $(\bar L_\alpha \gamma_\mu L_\beta)(\bar Q_\gamma \gamma^\mu Q_\delta)$ & 
$\calO_{\ell q,\alpha \beta \gamma \delta}^{(3)}$ &  $(\bar L _\alpha \gamma_\mu \tau^I L_\beta) (\bar Q_\gamma \gamma^\mu \tau^I Q_\delta)$ \\
$\calO_{\ell u,\alpha \beta \gamma \delta}$ & $(\bar L_{\alpha} \gamma_\mu L_{\beta}) (\bar u_\gamma \gamma^\mu u_\delta)$  &
$\calO_{\ell d,\alpha \beta \gamma \delta}$ &   $(\bar L_{\alpha} \gamma_\mu L_{\beta}) (\bar d_\gamma \gamma^\mu d_\delta)$ \\
$\calO_{eu,\alpha \beta \gamma \delta}$ &  $(\bar e_{\alpha} \gamma_\mu e_{\beta}) (\bar u_\gamma \gamma^\mu u_\delta)$ &
$\calO_{ed,\alpha \beta \gamma \delta}$ &   $(\bar e_{\alpha} \gamma_\mu e_{\beta}) (\bar d_\gamma \gamma^\mu d_\delta)$ \\
$\calO_{qe,\alpha \beta \gamma \delta}$ &   $(\bar Q_\alpha \gamma^\mu Q_\beta) (\bar e_{\gamma} \gamma_\mu e_{\delta})$ &
$\calO_{\ell edq,\alpha \beta \gamma \delta}$ &   $(\bar L_\alpha e_{\beta}) (\bar d_\gamma Q_\delta)$  \\
$\calO_{\ell equ,\alpha \beta \gamma \delta}^{(1)}$ &   $(\bar L_\alpha^a e_{\beta})\epsilon_{ab} (\bar Q_\gamma^b u_\delta)$ &
$\calO_{\ell equ,\alpha \beta \gamma \delta}^{(3)}$ &   $(\bar L_\alpha^a \sigma_{\mu\nu} e_{\beta})  \epsilon_{ab} (\bar Q_\gamma^b \sigma^{\mu\nu} u_\delta )$ \\[1ex]
\hline
\multicolumn{2}{c}{4$\ell$ operators} &
\multicolumn{2}{c}{Dipole operators} \\
\hline \\[-2.6ex]
$\calO_{\ell\ell,\alpha \beta \gamma \delta}$ & $ (\bar L _\alpha \gamma_\mu L_\beta) (\bar L_\gamma \gamma^\mu L_\delta)$ & 
$\calO_{eW,\alpha \beta}$ & $(\bar L_\alpha \sigma^{\mu\nu} e_{\beta}) \tau^I H W_{\mu\nu}^I$  \\
$\calO_{ee,\alpha \beta \gamma \delta}$ & $(\bar e_\alpha \gamma_\mu e_\beta) (\bar e_\gamma \gamma^\mu e_\delta)$ &
$\calO_{eB,\alpha \beta}$ & $(\bar L_\alpha \sigma^{\mu\nu} e_{\beta}) H B_{\mu\nu}$ \\
$\calO_{\ell e,\alpha \beta \gamma \delta}$ & $(\bar L_\alpha \gamma_\mu L_\beta) (\bar e_\gamma \gamma^\mu e_\delta)$ & & \\[.6ex]
\hline
\multicolumn{4}{c}{Lepton-Higgs operators}\\ 
\hline\\[-2.6ex]
$\calO_{H \ell,\alpha \beta}^{(1)}$ &  $(H^\dagger i \overleftrightarrow{D}_\mu H) ( \bar L_\alpha \gamma^\mu L_\beta)$ &
$\calO_{H \ell,\alpha \beta}^{(3)}$ &  $(H^\dagger i \overleftrightarrow{D}^I_\mu H) ( \bar L_\alpha \gamma^\mu \tau^I L_\beta)$ \\
$\calO_{H e, \alpha \beta}$ &  $(H^\dagger i \overleftrightarrow{D}_\mu H) ( \bar e_\alpha \gamma^\mu e_\beta)$ &
$\calO_{eH,\alpha \beta}$ &  $( \bar L_\alpha  e_\beta H) (H^\dagger H)$ \\[.6ex]
\hline
\end{tabular}
\caption{
Dimension-$6$ SMEFT operators relevant for cLFV processes, with flavour indices $\alpha, \beta, \gamma, \delta$.
$Q$ and $L$ are quark and lepton $SU(2)_L$ doublets ($a,b=1,2$, $I=1,2,3$ $SU(2)_L$ indices and $\tau^I$ the Pauli matrices), while $u$, $d$ and $e$ are up, down quark and lepton singlets. $H$ denotes the Higgs doublet with $\hj\equiv H^\dag (D_\mu H)-(D_\mu H)^\dag H$, and $B_{\mu\nu}$ and $W^I_{\mu\nu}$ are the $U(1)_Y$ and $SU(2)_L$ field strengths, respectively. 
\label{Tab:SMEFT}}
\end{table}

In the SMEFT, the SM is extended with a tower of new operators, each of them constructed with the SM particle content and respecting its symmetries, but suppressed with inverse powers of the scale $\Lambda$ at which the new degrees of freedom are expected,
\begin{equation}
\mathcal L_{\rm SMEFT} = \mathcal L_{\rm SM}  
+ \sum_{n>4}\frac1{\Lambda^{n-4}} C_{n} \mathcal O_{n}\,,
\end{equation}
with $C_n$ the Wilson coefficients (WCs) associated to each of the dim-$n$ effective operators.
At lowest order, cLFV observables are introduced at dimension 6 by dipole, lepton-Higgs and 4-fermion operators, as summarized in Table~\ref{Tab:SMEFT} in the Warsaw basis~\cite{Grzadkowski:2010es}.

On the other hand, the most relevant observables to constrain charged lepton flavour change are decays of light mesons and charged leptons. As such, the appropriate EFT to describe these low-energy processes is the LEFT~\cite{Jenkins:2017jig}, sometimes also referred to as WEFT or WET, built from the particle content of the SM minus the heavy degrees of freedom (top, Higgs, $W$ and $Z$) and respecting the unbroken QED and QCD symmetries. 
The lowest order cLFV operator in the LEFT is the dimension-5 dipole operator, 
\begin{equation}\label{eq:LEFTdim5}
\mathcal L_{\rm LEFT}^{\rm dim-5} \supset 
\frac{\sqrt{2}}{v}\, \sum_{\alpha\neq\beta}\,  \wcL{\alpha\beta}{e\gamma}\,(\bar e_{L_\alpha} \sigma^{\mu\nu} e_{R_\beta}) F_{\mu\nu} + \hc
\end{equation}
At dimension 6, the operators of interest would be 4-fermion operators containing (at least) two charged leptons, 
\begin{equation}
\mathcal L_{\rm LEFT}^{\rm dim-6} \supset 
\frac2{v^2} \sum_{q,x,Y}\, \wcL{\alpha\beta X}{qx}\,\mathcal O_{\alpha\beta X}^{qx} +
\frac2{v^2} \sum_{y,X,Y}\, \wcL{\alpha\beta X}{\gamma\delta Yy }\,\mathcal O_{\alpha\beta X}^{\gamma\delta Yy}\,,
\end{equation}
where $v=(\sqrt2 G_F)^{-1/2}$, with $G_F$ the Fermi constant, and the sums run over the quarks $q$, the lepton chiralities $X,Y=L,R$ and the Lorentz structures $x=V,A, S, P, T$ for quarks and $y=V,S$ for leptons.
The lepton flavour indices $\alpha,\beta,\gamma,\delta=e,\mu,\tau$ are again assumed to be LFV (at least) in the $\alpha$-$\beta$ sector, while the quark sector will be assumed to be flavour diagonal and composed only of $u$, $d$ and $s$ quarks, as discussed in the introduction. We will also assume all WCs to be real.

\begin{table}[t!]

%
\mbox{}\\[-1.25cm]

\begin{adjustbox}{width=1.05\textwidth,center}
\begin{minipage}[t]{3cm}
\renewcommand{\arraystretch}{1.51}
\small
\begin{align*}
\begin{array}[t]{c|c}
\multicolumn{2}{c}{\bold{Leptonic}} \\
\hline
\opd{\alpha \beta L}{\gamma \delta LV}       & (\bar e_{L \alpha}  \gamma^\mu e_{L \beta})(\bar e_{L \gamma} \gamma_\mu e_{L\delta})   \\
\opd{\alpha \beta R}{\gamma \delta RV}    & (\bar e_{R \alpha} \gamma^\mu e_{R\beta})(\bar e_{R\gamma} \gamma_\mu e_{R\delta})  \\
\opd{\alpha \beta L}{\gamma \delta RV}       & (\bar e_{L \alpha}  \gamma^\mu e_{L\beta})(\bar e_{R\gamma} \gamma_\mu e_{R\delta}) \\
\opd{\alpha \beta R}{\gamma \delta LV}       & (\bar e_{R \alpha}  \gamma^\mu e_{R\beta})(\bar e_{L\gamma} \gamma_\mu e_{L\delta}) \\
\hline
\opd{\alpha \beta R}{\gamma \delta RS}		& (\bar e_{L \alpha}   e_{R\beta}) (\bar e_{L\gamma} e_{R \delta})  +\hc \\
\multicolumn{2}{c}{\bold{Dipole}} \\
\hline
\opd{\alpha \beta} {e \gamma}     & (\bar e_{L \alpha} \sigma^{\mu\nu} e_{R\beta}) F_{\mu\nu}  +\hc  \\
\end{array}
\end{align*}
\end{minipage}
\begin{minipage}[t]{3cm}
\renewcommand{\arraystretch}{1.51}
\small
\begin{align*}
\begin{array}[t]{c|c}
\multicolumn{2}{c}{\bold{up-quarks}} \\
\hline
\opd{\alpha \beta L}{uV}   &  (\bar u \gamma_\mu u)(\bar e_{L \alpha}  \gamma^\mu e_{L\beta})   \\
\opd{\alpha \beta L}{uA}    & (\bar u \gamma_\mu \gamma_5 u)(\bar e_{L \alpha}  \gamma^\mu e_{L\beta})   \\
\opd{\alpha \beta R}{uV}      & (\bar u \gamma_\mu u)(\bar e_{R \alpha}  \gamma^\mu e_{R\beta})   \\
\opd{\alpha \beta R}{uA}    & (\bar u \gamma_\mu \gamma_5 u)(\bar e_{R \alpha}  \gamma^\mu e_{R\beta})   \\
\hline
\opd{\alpha \beta R}{uS}  & (\bar u  u)(\bar e_{L \alpha} e_{R\beta})   +\hc \\
\opd{\alpha \beta R}{uP}  & (\bar u \gamma_5 u)(\bar e_{L \alpha}   e_{R\beta})   +\hc \\
\opd{\alpha \beta R}{uT} &  (\bar u  \sigma_{\mu \nu}  u)(\bar e_{L \alpha}   \sigma^{\mu \nu}   e_{R\beta})   +\hc \\
\end{array}
\end{align*}
\end{minipage}
\begin{minipage}[t]{3cm}
\renewcommand{\arraystretch}{1.51}
\small
\begin{align*}
\begin{array}[t]{c|c}
\multicolumn{2}{c}{\bold{down-quarks}} \\
\hline
\opd{\alpha \beta L}{dV}   &  (\bar d \gamma_\mu d)(\bar e_{L \alpha}  \gamma^\mu e_{L\beta})   \\
\opd{\alpha \beta L}{dA}    & (\bar d \gamma_\mu \gamma_5 d)(\bar e_{L \alpha}  \gamma^\mu e_{L\beta})   \\
\opd{\alpha \beta R}{dV}      & (\bar d \gamma_\mu d)(\bar e_{R \alpha}  \gamma^\mu e_{R\beta})   \\
\opd{\alpha \beta R}{dA}    & (\bar d \gamma_\mu \gamma_5 d)(\bar e_{R \alpha}  \gamma^\mu e_{R\beta})   \\
\hline
\opd{\alpha \beta R}{dS}  & (\bar d  d)(\bar e_{L \alpha} e_{R\beta})   +\hc \\
\opd{\alpha \beta R}{dP}  & (\bar d \gamma_5 d)(\bar e_{L \alpha}   e_{R\beta})   +\hc \\
\opd{\alpha \beta R}{dT} &  (\bar d  \sigma_{\mu \nu}  d)(\bar e_{L \alpha}   \sigma^{\mu \nu}   e_{R\beta})   +\hc \\
\end{array}
\end{align*}
\end{minipage}

\end{adjustbox}
\setlength{\abovecaptionskip}{0.15cm}
\caption{The list of relevant $d=6$ LEFT operators for processes with charged lepton flavour violation ($\alpha \neq \beta$) extracted from Ref.~\cite{Jenkins:2017jig}.
Also included the $d=5$ dipole operator.}
\label{tab:LEFT}
\end{table}

The operators of our LEFT basis are listed in Table~\ref{tab:LEFT}, taken from Ref.~\cite{Jenkins:2017jig} but with small modifications\footnote{Notice that we consider $\op{\alpha \beta L}{V}{\gamma \delta R} $ and  $\op{\alpha \beta R}{V}{\gamma \delta L} = \op{\gamma \delta L}{V}{\alpha \beta R} $ as two different operators, since the former/latter introduces a left/right cLFV current. This will be relevant, for instance, when matching to the SMEFT.}.
In particular, we choose a basis with definite chirality for the charged leptons, since chirality flips are suppressed by their mass and can be largely neglected, but separating vector from axial and scalar from pseudoscalar structures for the quark bilinears, since in this way it is most straightforward to connect each operator with decays of and into pseudoscalar or vector mesons, as well as with spin independent or dependent contributions to $\mu - e$ conversion in nuclei.

When these LEFT operators are matched to the SMEFT, the SU(2)$_L$ structure imposes non-trivial correlations and constraints altering the counting of relevant independent operators~\cite{Jenkins:2017jig}. In particular, the scalar and tensor structures are much simpler, since several of them are not generated by the SMEFT at $d=6$ (they are QED invariant, but not $U(1)_Y$ invariant), and those that are generated match to a single operator\footnote{Notice that, these relations will hold matching the two theories at the electroweak scale. We neglect the modifications that may come from running down to the low scale relevant for the observables.}. More precisely, 
\begin{align}
%
%
&\wcL{\alpha\beta R}{\gamma\delta R S} = 0\,, \label{eq:smefttoleftfisrt} \\
&\wcL{\alpha\beta R}{u S} = \wcL{\alpha\beta R}{u P} = -\frac{v^2}{2\Lambda^2}\,C_{\ell equ}^{(1)\,\alpha\beta uu}\,,&
\wcL{\alpha\beta R}{u T} &= -\frac{v^2}{2\Lambda^2}\,C_{\ell equ}^{(3)\,\alpha\beta uu}\,, \\
&\wcL{\alpha\beta R}{d S} = -\wcL{\alpha\beta R}{d P} = \frac{v^2}{2\Lambda^2}\,C_{\ell edq}^{\alpha\beta uu}\,,&
\wcL{\alpha\beta R}{d T} &= 0\,.
\label{eq:smefttoleft}
\end{align}
Consequently, the scalar and tensorial LEFT sector is much more general than that of the SMEFT. In Section~\ref{sec:SMEFT}, we will make use of this feature to further restrict our global analysis after presenting the status for the LEFT framework.

Conversely, and in contrast to the scalar and tensor operators, the 4-fermion vector and axial operators generally receive contributions from two SMEFT counterparts. Indeed, not only the 4-fermion SMEFT operators coupling two lepton currents are present, but also operators inducing a coupling between a (flavour changing) lepton current and the $Z$ boson after the latter is integrated out. In particular, the combination  $C^{(1)}_{H\ell}+C^{(3)}_{H\ell}$ for left-handed currents and $\wcH{He}{}$ for right-handed. All in all, keeping only the leading SMEFT terms, the tree-level matching reads:  
\begin{align}
%
%
\label{eq:match1}
\tfrac{2\Lambda^2}{v^2}\,\wcL{\alpha\beta L}{\gamma\delta L V} &= 
\wcH{\ell\ell}{\alpha\beta\gamma\delta} - \tfrac14 \left(1-2\sw^2 \right)\delta_{\gamma\delta}\big(1+\delta_{\alpha\gamma}+\delta_{\beta\gamma}\big)\Big[\wcH{H\ell}{(1)}+\wcH{H\ell}{(3)}\Big]^{\alpha\beta},\\
\tfrac{2\Lambda^2}{v^2}\,\wcL{\alpha\beta R}{\gamma\delta R V} &= 
\wcH{ee}{\alpha\beta\gamma\delta} + \tfrac12\sw^2\,\delta_{\gamma\delta}\big(1+\delta_{\alpha\gamma}+\delta_{\beta\gamma}\big)\,\wcH{He}{\alpha\beta}\,,\\
\tfrac{2\Lambda^2}{v^2}\,\wcL{\alpha\beta L}{\gamma\delta R V} &= 
\wcH{\ell e}{\alpha\beta\gamma\delta} + 2\sw^2\,\delta_{\gamma\delta}\Big[\wcH{H\ell}{(1)}+\wcH{H\ell}{(3)}\Big]^{\alpha\beta},\\
\tfrac{2\Lambda^2}{v^2}\,\wcL{\alpha\beta R}{\gamma\delta L V} &= 
\wcH{\ell e}{\gamma\delta\alpha\beta} - \left(1-2\sw^2 \right)\delta_{\gamma\delta}\,\wcH{He}{\alpha\beta}\,,\\
%
%
\tfrac{2\Lambda^2}{v^2}\,\wcL{\alpha\beta L}{u V} &=  
\tfrac12\Big[\wcH{\ell u}{}+\wcH{\ell q}{(1)}-\wcH{\ell q}{(3)}\Big]^{\alpha\beta uu} 
+ \Big(\tfrac12-\tfrac43\sw^2\Big)\Big[\wcH{H\ell}{(1)}+\wcH{H\ell}{(3)}\Big]^{\alpha\beta},\\
\tfrac{2\Lambda^2}{v^2}\,\wcL{\alpha\beta L}{u A} &=  
\tfrac12\Big[\wcH{\ell u}{}-\wcH{\ell q}{(1)}+\wcH{\ell q}{(3)}\Big]^{\alpha\beta uu} 
-   \tfrac12\,\Big[\wcH{H\ell}{(1)}+\wcH{H\ell}{(3)}\Big]^{\alpha\beta},\\
\tfrac{2\Lambda^2}{v^2}\,\wcL{\alpha\beta R}{u V} &=  
\tfrac12\Big[\wcH{eu}{}+\wcH{qe}{}\Big]^{\alpha\beta uu} 
+ \Big(\tfrac12-\tfrac43\sw^2\Big)\,\wcH{He}{\alpha\beta}\,,\\
\tfrac{2\Lambda^2}{v^2}\,\wcL{\alpha\beta R}{u A} &=  
\tfrac12\Big[\wcH{eu}{}-\wcH{qe}{}\Big]^{\alpha\beta uu} 
-\tfrac12\,\wcH{He}{\alpha\beta}\,,\\
%
%
\tfrac{2\Lambda^2}{v^2}\,\wcL{\alpha\beta L}{d V} &= 
\tfrac12\Big[\wcH{\ell d}{}+\wcH{\ell q}{(1)}+ \wcH{\ell q}{(3)}\Big]^{\alpha\beta dd} 
-\Big(\tfrac12-\tfrac23\sw^2\Big)\Big[\wcH{H\ell}{(1)}+\wcH{H\ell}{(3)}\Big]^{\alpha\beta},\\
\tfrac{2\Lambda^2}{v^2}\,\wcL{\alpha\beta L}{d A} &= 
\tfrac12\Big[\wcH{\ell d}{}-\wcH{\ell q}{(1)}- \wcH{\ell q}{(3)}\Big]^{\alpha\beta dd} 
+ \tfrac12\,\Big[\wcH{H\ell}{(1)}+\wcH{H\ell}{(3)}\Big]^{\alpha\beta},\\
\tfrac{2\Lambda^2}{v^2}\,\wcL{\alpha\beta R}{d V} &=  
\tfrac12\Big[\wcH{ed}{}+\wcH{qe}{}\Big]^{\alpha\beta dd} 
- \Big(\tfrac12-\tfrac23\sw^2\Big)\,\wcH{He}{\alpha\beta}\,,\\
\label{eq:matchlast}
\tfrac{2\Lambda^2}{v^2}\,\wcL{\alpha\beta R}{d A} &= 
\tfrac12\Big[\wcH{ed}{}-\wcH{qe}{}\Big]^{\alpha\beta dd} 
+ \tfrac12\,\wcH{He}{\alpha\beta}\,, 
\end{align}
where, for simplicity and since we are focusing on the flavour-change in the lepton sector, we have set the CKM matrix to the identity. Notice that the effect of the CKM would be to include additional SMEFT operators involving the $s$ and $b$ quarks together with the $d$ providing a (CKM-suppressed) contribution to the $d$ mass eigenstate. Since these operators would induce flavour violation both in the quark and lepton sectors, they would be constrained with additional observables. 

From the above relations, we see that 14 different SMEFT operators (although in 13 independent combinations) match into only 12 LEFT operators. This means that low-energy cLFV observables cannot be enough to fully constrain the vectorial sector of the SMEFT. Fortunately, higher energy observables where the $Z$ is not integrated out allow to disentangle among the several SMEFT operators that contribute to a given LEFT one.

In particular, $C_{H\ell}\equiv C^{(1)}_{H\ell}+C^{(3)}_{H\ell}$ and $C_{He}$ will induce cLFV decays of an on-shell $Z$-boson. These decays have been searched for and are strongly constrained by the LHC~\cite{ATLAS:2022uhq,ATLAS:2021bdj} and directly probe the operators of interest~\cite{Crivellin:2013hpa,Calibbi:2021pyh}:
\begin{equation}
    \hspace{-.32cm}
    \text{BR}\big(Z\rightarrow \ell_{\alpha}^{-}\ell_{\beta}^{+}\big)=\dfrac{G_F M_Z^3}{3\sqrt{2} \pi\Gamma_Z}
    \hspace{-.1cm}
    \left\{\abs{\frac{v^2}{2\Lambda^2} C^{\alpha\beta}_{H\ell}}^2+\abs{\frac{v^2}{2\Lambda^2} C^{\alpha\beta}_{He}}^2 + \dfrac{1}{4}\abs{\frac{v^2}{2\Lambda^2} C^{\alpha\beta}_{eZ}}^2+\dfrac{1}{4}\abs{\frac{v^2}{2\Lambda^2} C^{\beta\alpha}_{eZ}}^2\right\},
    \label{eq:Zdecay}
\end{equation}
where we have defined 
\begin{equation}
C^{\alpha\beta}_{eZ}\equiv \Big[\sw \wcH{eB}{} + \cw \wcH{eW}{}\Big]^{\alpha\beta}\,.
\end{equation}
Since both vertex corrections contribute incoherently, it is possible to extract bounds on each of the WCs, which in turn would allow to disentangle and constrain the SMEFT 4-fermion operators in the r.h.s.~of Eqs.~\eqref{eq:match1} to~\eqref{eq:matchlast}, provided the corresponding LEFT coefficient in the l.h.s.~is bounded.

Finally, and for completeness, the matching for the dipole operator is given by:
\begin{equation}
\wcL{\alpha\beta}{e\gamma} =  \frac{v^2}{2 \Lambda^2} \Big[\cw \wcH{eB}{} - \sw \wcH{eW}{}\Big]^{\alpha\beta}\,.
\end{equation}

Thus, assuming that LEFT operators originate from the low-energy contribution of the $d=6$ SMEFT implies, overall, a significant reduction on the amount of free parameters given the strong restrictions imposed in the scalar and tensor sectors. In Section~\ref{sec:LEFT} we will derive general model-independent constraints through the LEFT paradigm and then, in Section~\ref{sec:SMEFT}, we will restrict our study to the subset of LEFT operators generated from the low-energy $d=6$ SMEFT. Disentangling among the different SMEFT contributions would simply require to consider the constraints from LFV $Z$ decays on the corresponding operators (see Eq.~\eqref{eq:Zdecay}) and then run down and match to the LEFT to combine with the low-energy bounds we will derive in this work and present in the following sections.

\section{Radiative and three-body cLFV decays}
\label{sec:lepobs}

\begin{table}[t!]
\begin{center}
{\small
\renewcommand{\arraystretch}{1.3}
\begin{tabular}{lcll}
\hline
cLFV obs. & \multicolumn{2}{l}{Present upper bounds  $(90\%\,\text{CL})$} 
\\
\hline
BR$(\mu\to e\gamma)$ &  $3.1\times10^{-13}$ &MEG II (2023)&\cite{MEGII:2023ltw} 
\\
BR$(\mu\to eee)$ &   $1.0\times10^{-12}$ & SINDRUM (1988)&\cite{SINDRUM:1987nra} 
\\
CR$(\mu\to e,{\rm S})$ & $7.0\times10^{-11}$ &
Badertscher {\it et al.} (1982)&\cite{Badertscher:1981ay}
\\
CR$(\mu\to e,{\rm Ti})$ & $4.3\times10^{-12}$ &SINDRUM II (1993)&\cite{SINDRUMII:1993gxf}
\\
CR$(\mu\to e,{\rm Pb})$ & $4.6\times10^{-11}$ &SINDRUM II (1996)&\cite{SINDRUMII:1996fti}
\\
CR$(\mu\to e,{\rm Au})$ & $7.0\times10^{-13}$ &SINDRUM II (2006)&\cite{SINDRUMII:2006dvw}
\\
BR$(\pi^{0}\to \mu^{-} e^{+})$ &  $3.2\times10^{-10}$ & NA62 (2021)&\cite{NA62:2021zxl} 
\\
BR$(\pi^{0}\to \mu^{+} e^{-})$ &  $3.8\times10^{-10}$ & E865 (2000)&\cite{Appel:2000wg} 
\\
BR$(\pi^{0}\to \mu e)$ &  $3.6\times10^{-10}$ & KTeV (2007)&\cite{KTeV:2007cvy} 
\\
BR$(\eta\to \mu e)$ &  $6.0\times10^{-6}$ & Saturne SPES2 (1996)&\cite{White:1995jc} 
\\
BR$(\eta'\to \mu e)$ &  $4.7\times10^{-4}$ & CLEO (2000)&\cite{CLEO:1999nsy} 
\\
BR$(\phi\to \mu e)$ &  $2.0\times10^{-6}$ & SND (2009)&\cite{Achasov:2009en} 
\\
\hline
BR$(\tau\to e\gamma)$ & $3.3\times10^{-8}$ &BaBar (2010)&\cite{BaBar:2009hkt} 
\\
BR$(\tau\to ee\bar{e})$ & $2.7\times10^{-8}$ &Belle (2010)&\cite{Hayasaka:2010np} 
\\
BR$(\tau\to e\mu\bar{\mu})$ & $2.7\times10^{-8}$ &Belle (2010)&\cite{Hayasaka:2010np} 
\\
BR$(\tau\to e \pi)$ & $8.0\times10^{-8}$  &  Belle (2007)&\cite{Belle:2007cio} 
\\
BR$(\tau\to e \eta)$ & $9.2\times10^{-8}$  &  Belle (2007)&\cite{Belle:2007cio} 
\\
BR$(\tau\to e \eta')$ & $1.6\times10^{-7}$  &  Belle (2007)&\cite{Belle:2007cio}
\\
BR$(\tau\to e \pi\pi)$ & $2.3\times10^{-8}$ &Belle (2012)&\cite{Belle:2012unr}
\\
BR$(\tau\to e\omega)$ & $2.4\times10^{-8}$ &Belle (2023)&\cite{Belle:2023ziz} 
\\
BR$(\tau\to e\phi)$ & $2.0\times10^{-8}$ &Belle (2023)&\cite{Belle:2023ziz} 
\\
\hline
BR$(\tau\to \mu\gamma)$ & $4.2\times10^{-8}$ &Belle (2021)&\cite{Belle:2021ysv} 
\\
BR$(\tau\to \mu\mu\bar{\mu})$ & $2.1\times10^{-8}$ &Belle (2010)&\cite{Hayasaka:2010np} 
\\
BR$(\tau\to \mu e\bar{e})$ & $1.8\times10^{-8}$ &Belle (2010)&\cite{Hayasaka:2010np} 
\\
BR$(\tau\to \mu \pi)$ &  $1.1\times10^{-7}$ & BaBar (2006)&\cite{BaBar:2006jhm} 
\\
BR$(\tau\to \mu \eta)$ & $6.5\times10^{-8}$  &  Belle (2007)&\cite{Belle:2007cio} 
\\
BR$(\tau\to \mu \eta')$ & $1.3\times10^{-7}$  &  Belle (2007)&\cite{Belle:2007cio}
\\
BR$(\tau\to \mu\pi \pi)$ & $2.1\times10^{-8}$ &Belle (2012)&\cite{Belle:2012unr}
\\
BR$(\tau\to \mu\omega)$ & $3.9\times10^{-8}$ &Belle (2023)&\cite{Belle:2023ziz} 
\\
BR$(\tau\to \mu\phi)$ & $2.3\times10^{-8}$ &Belle (2023)&\cite{Belle:2023ziz} 
\\
\hline
BR$(\tau\to ee\bar{\mu})$ & $1.5\times10^{-8}$ &Belle (2010)&\cite{Hayasaka:2010np}
\\
BR$(\tau\to \mu\mu\bar{e})$ & $1.7\times10^{-8}$ &Belle (2010)&\cite{Hayasaka:2010np}
\\
\hline
\end{tabular}
\caption{Set of low-energy cLFV current bounds relevant for our analysis.}
}
\end{center}
\label{Tab:LFVlimits}
\end{table}

Dipole and 4-lepton operators can induce cLFV radiative decays $\ell_\alpha\rightarrow\ell_\beta\gamma$ and three body decays $\ell_\alpha\rightarrow \ell_\beta \ell_\gamma\bar{\ell}_\delta$.
See Table~\ref{Tab:LFVlimits} for the limits on these processes.
The expressions for their respective branching ratios can be found in the literature, see for instance Refs.~\cite{Kuno:1999jp,Brignole:2004ah,Crivellin:2017rmk,Calibbi:2021pyh}. Therefore, we will refrain from explicitly showing them, as the crucial observation to be made is that each operator contributes incoherently. This implies that the extraction of bounds on the WCs under the assumption that only one operator is present at a time is equivalent to performing a global fit.
We will thus present the bounds on the cLFV LEFT operators that can be constrained with the aforementioned low-energy observables, while also listing the operators that are unconstrained. 

\begin{figure}[t!]
    \centering
    \includegraphics[width=0.9\textwidth]{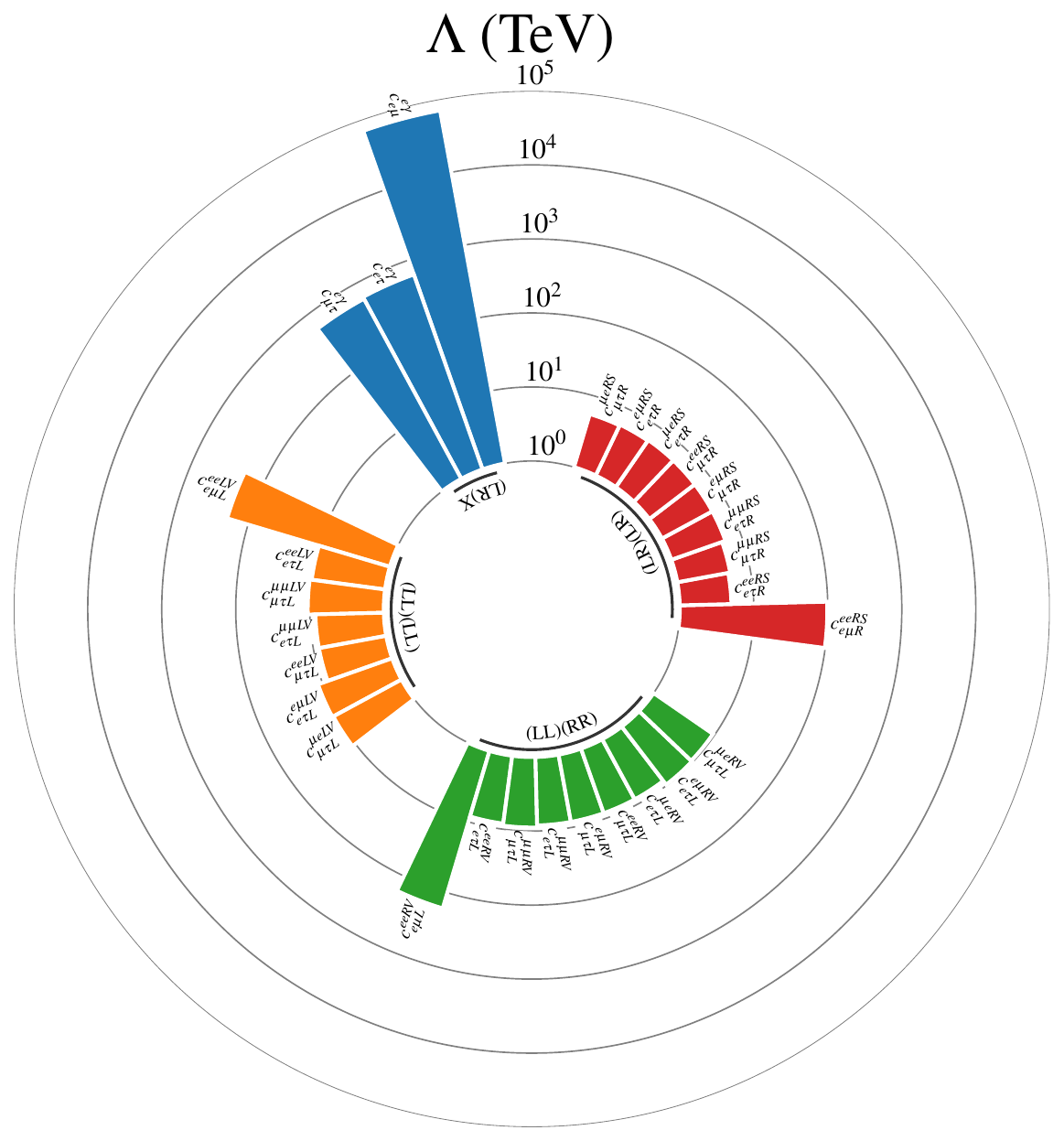}
    \caption{Current 95\% CL upper bounds on 4-lepton and dipole LEFT operators. The same bounds hold for the corresponding $(L \longleftrightarrow R)$ operators. In the case in which the LEFT is matched as the low-energy realisation of the SMEFT, the bounds on scalar operators do not apply, since they are not generated, but the rest still hold. The bounds shown in the plot are collected in appendix~\ref{app:dipole_4l_bounds}.}
    \label{fig:circular_barplot}
\end{figure}

The bounds on the WCs are shown in Fig.~\ref{fig:circular_barplot}.
All cLFV dipole operators can be simultaneosuly constrained by the three radiative decays. This family of operators can also mediate three-body decays and other cLFV processes we will discuss later, such as $\mu-e$ conversion in nuclei. However, due to the very stringent bounds imposed by the radiative decays, we will neglect them for the rest of our discussion.

As for cLFV 4-lepton operators, the situation is slightly more involved and depends on the flavour structure of the operators.
Considering for instance the Lorentz structure with two left-handed currents and for $\alpha\neq\beta$, there are 15 independent operators. Out of those, 7 are constrained by three body decays, leaving 8 operators unbounded. The exact same counting holds for the operators with two right-handed currents. In the case of the vector structures coupling $L$ and $R$ currents, out of the 33 operators, 18 can be constrained, leaving 15 operators without bound. Finally, for scalar structures 18 out of 36 are bounded and thus 18 remain unconstrained. 

The flavour structures of the unconstrained 4-lepton operators involve flavour combinations in which a decay is kinematically forbidden, since the heaviest lepton field involved in the operator appears several times, and thus are much less straightforward to probe. 
In particular, there are 8 such flavour combinations, which can violate flavour in either one or two units,  
\begin{align}
\label{eq:dF1}
    \bar e\mu \bar\mu\mu,~ \bar e\tau\bar\tau\tau,~ \bar \mu\tau\bar\tau\tau,~ \bar e\mu\bar\tau\tau &\qquad (\Delta F = 1)\,,\\
    \bar e\mu \bar e\mu,~ \bar e\tau\bar e\tau,~ \bar\mu\tau\bar\mu\tau,~\bar e\tau\bar\mu\tau &\qquad (\Delta F = 2)\,,
\end{align}
and come in all the possible quiralities and Lorentz structures\footnote{For vectorial operators with two left or right currents, $c^{\gamma\delta XV}_{\alpha\beta X}$, these 8 are the only independent flavour combinations. For the $c^{\gamma\delta LV}_{\alpha\beta R}$ and $c^{\gamma\delta RS}_{\alpha\beta R}$ structures, however, there are additional ones which cannot be related to the above via Fierz identities, explaining the difference in the previous counting.} given in Table~\ref{tab:LEFT}. 
Interestingly, the operators with 2 muons and 2 electrons, which simultaneously violate $L_\mu$ and $L_e$ in 2 units, are not entirely unconstrained, since they induce muonium-antimuonium oscillations $M_\mu\to\bar{M_\mu}$. The rather strong constraints on this process~\cite{PhysRevLett.82.49} can be used to extract bounds on the corresponding operators~\cite{Conlin:2020veq} (see also Ref.~\cite{Heeck:2024uiz} for a recent discussion on bounds for $\Delta F=2$ operators). Furthermore, operators with $\Delta F = 1$ in Eq.~\eqref{eq:dF1} will mix with other $\Delta F = 1$ operators for which bounds do exist (such as the $\bar \mu e \bar e e$ structure). Nevertheless, we do not include this effect as in this work we always consider all WCs directly at the low scale relevant for the observables under study.

All in all, no global fit is required for dipole and 4-lepton operators due to their incoherent contributions. Consequently, the LEFT bounds displayed in Fig.~\ref{fig:circular_barplot} can be regarded as global. These bounds also directly apply to the scenario where the LEFT operators arise as the low-energy operators of the $d=6$ SMEFT, which could be then translated to the SMEFT WCs using Eqs.~\eqref{eq:smefttoleftfisrt} to~\eqref{eq:matchlast}, with the caveat of those operators which receive contributions from the LFV coupling to the $Z$. In these instances, the bounds should be combined with LFV $Z$ decays, as outlined above, to independently constrain the different contributions.

\section{Semileptonic cLFV observables}
\label{sec:obs}

We now shift the focus to cLFV $2\ell2q$ operators, which induce, among other processes, $\mu-e$ conversion in nuclei, leptonic meson decays and semileptonic tau decays.
Contrary to the previous section, the $2\ell2q$ operators contribute in different coherent combinations and thus lead to a non-trivial analysis of constrained and flat directions.
This will result in potentially different bounds depending on whether the approach is global or assumes only one operator at a time, as we will quantify later in sections~\ref{sec:LEFT}, \ref{sec:SMEFT} and~\ref{sec:SMEFT-s}. 
In this section, we start presenting all relevant expressions for our analysis, given in terms of the LEFT operators of Table~\ref{tab:LEFT} at the relevant scale for each observable. 

\subsection[cLFV semileptonic $\tau$ decays]{cLFV semileptonic $\boldsymbol{\tau}$ decays}
\label{subsec:tauLFVobservables}
Operators involving flavour change between a $\tau$ and a lighter charged lepton with two quarks induce decays of the $\tau$ into the lighter charged lepton and one or two mesons. $B$ factories are ideal to probe for these rare decays and stringent constraints exist for their branching ratios from BaBar and Belle (see Table~\ref{Tab:LFVlimits}).

Although the number of WCs to probe is high and it involves different chiral structures (see Table~\ref{tab:LEFT}), the plethora of meson final states listed in Table~\ref{Tab:LFVlimits} with available bounds offers great complementarity, allowing to constrain many directions in parameter space. For example, the decay $\tau\rightarrow\ell \pi^0$ provides sensitivity to the axial and pseudoscalar WCs for both chiralities of the charged lepton~\cite{Aebischer:2018iyb}: 
\begin{equation}
    \hspace{-.24cm}
    \text{BR}\left(\tau\rightarrow \ell\pi^0\right)=\dfrac{G_F^2 f_\pi^2 (m_\tau^2-m_\pi^2)^2}{8\pi \Gamma_\tau m_\tau}
    \hspace{-.25cm}
    \sum_{X=L,R}\left\lvert c^{uA}_{\tau\ell X}-c^{dA}_{\tau\ell X}+\dfrac{m_\pi^2}{m_\tau\left(m_u+m_d \right)}\left(c_{\tau\ell X}^{uP}-c_{\tau\ell X}^{dP}\right)\right\rvert^2,
\label{eq:tau2pi0}
\end{equation}
whereas the decay $\tau\rightarrow \ell \pi^+\pi^-$ gives access to the vector/tensor and scalar WCs via the $\rho$ and $f^0$ resonances, respectively. The computation of the latter requires the input of different hadronic matrix elements (see {\it e.g.}~\cite{Celis:2014asa,Cirigliano:2021img}). Neglecting interference terms, the parametric dependence of the BR in terms of the WCs is~\cite{Cirigliano:2021img}:
\begin{align}
    \text{BR}\left(\tau\rightarrow \ell\pi^+\pi^-\right)=\sum_{X=L,R}\Bigg\{
    2.0&\left\lvert c^{uV}_{\tau\ell X} - c^{dV}_{\tau\ell X}\right\rvert^2
    +0.68\left\lvert c^{uS}_{\tau\ell X} +c^{dS}_{\tau\ell X}\right\rvert^2
    +0.52\left\lvert c^{sS}_{\tau\ell X}\right\rvert^2\nonumber\\
    +4.0&\left\lvert c^{uT}_{\tau\ell X} - c^{dT}_{\tau\ell X}\right\rvert^2\Bigg\}\,.
    \label{eq:tau2pipi}
\end{align}

In addition to the spin of the final meson, different isospin\footnote{Since exact isospin symmetry is assumed for the estimation of matrix elements for the $\tau\rightarrow \ell \pi^+\pi^-$ decay, we assumed it for all observables involving $\tau$-s so as to avoid artificially lifting potential flat directions.} structures also allows to constrain different combinations of WCs. For instance, while decays into the $\rho$-resonance allow to constrain the isovector combination of vector coefficients, the $\tau\rightarrow\ell\omega$ decay can be used to constrain the isoscalar combination\footnote{It should be noted that there is an interference term between the vector and tensor operators that is not chirally-suppressed. However, since this term does not lead to coherent interference, we do not include it, as it would only lead to a minor modification of our results.} instead~\cite{Aebischer:2018iyb} 
\begin{align}
    \text{BR}\left(\tau\rightarrow \ell \omega\right)&=\dfrac{G_F^2 f_{\omega}^2m_{\omega}^2 (m_\tau^2-m_\omega^2)}{8\pi \Gamma_\tau m_\tau}\sum_{X=L,R}\bigg\{ \left(\frac{m_\tau^2}{m_\omega^2}+1-2\frac{m_\omega^2}{m_\tau^2}\right)\left\lvert c^{uV}_{\tau\ell X} + c^{dV}_{\tau\ell X}\right\rvert^2 \nonumber\\
    &+4\left(\frac{ f_{T,\omega}}{f_\omega}\right)^2\left(2\frac{m_\tau^2}{m_\omega^2}-1-\frac{m_\omega^2}{m_\tau^2}\right)\left\lvert c^{uT}_{\tau\ell X}+c^{dT}_{\tau\ell X}\right\rvert^2 \bigg\}.
    \label{eq:tau2omega}
\end{align}

Similarly, the $\tau\rightarrow\ell\eta$ and $\tau\rightarrow\ell\eta'$ decays~\cite{Celis:2014asa}  constrain the isoscalar and $s$-quark combinations of pseudoscalar and axial operators, complementing the constrain from $\tau\rightarrow\ell\pi^0$:
\begin{align}
    \text{BR}\left(\tau\rightarrow \ell\eta\right)=\dfrac{G_F^2 f_\pi^2 \big(m_\tau^2-m_\eta^2\big)^2}{8\pi \Gamma_\tau m_\tau}&\sum_{X=L,R}\Bigg\lvert 
    \dfrac{f_\eta^{u}}{f_\pi}\Big(c^{uA}_{\tau\ell X}+c^{dA}_{\tau\ell X}\Big)
    + \dfrac{f_\eta^{s}}{f_\pi}c^{sA}_{\tau\ell X}\nonumber\\
    +&\dfrac{h_\eta^{u}}{f_\pi m_\tau(m_u+m_d)}\Big(c^{uP}_{\tau\ell X}+c_{\tau\ell X}^{dP}\Big)+\dfrac{h_\eta^{s}}{2f_\pi m_\tau m_s}c^{sP}_{\tau\ell X}\Bigg\rvert^2,
    \label{eq:tau2eta}
\end{align}
\begin{align}
    \text{BR}\left(\tau\rightarrow \ell\eta'\right)=\dfrac{G_F^2 f_\pi^2 \big(m_\tau^2-m_{\eta'}^2\big)^2}{8\pi \Gamma_\tau m_\tau}&\sum_{X=L,R}\Bigg\lvert 
    \dfrac{f_{\eta'}^{u}}{f_\pi}\Big(c^{uA}_{\tau\ell X}+c^{dA}_{\tau\ell X}\Big)
    + \dfrac{f_{\eta'}^{s}}{f_\pi}c^{sA}_{\tau\ell X}\nonumber\\
    +&\dfrac{h_{\eta'}^{u}}{f_\pi m_\tau(m_u+m_d)}\Big(c^{uP}_{\tau\ell X}+c_{\tau\ell X}^{dP}\Big)+\dfrac{h_{\eta'}^{s}}{2f_\pi m_\tau m_s}c^{sP}_{\tau\ell X}\Bigg\rvert^2.
\label{eq:tau2etaprime}
\end{align}
Finally, the decay $\tau\rightarrow \ell\phi$ can be used to access the vector and tensor involving $s$-quarks~\cite{Aebischer:2018iyb}:
\begin{align}
    \text{BR}\left(\tau\rightarrow \ell \phi\right)=\dfrac{G_F^2 f_{\phi}^2m_{\phi}^2 (m_\tau^2-m_\phi^2)}{4\pi \Gamma_\tau m_\tau}\sum_{X=L,R}\Bigg\{& \left(\frac{m_\tau^2}{m_\phi^2}+1-2\frac{m_\phi^2}{m_\tau^2}\right)\left\lvert c^{sV}_{\tau\ell X}\right\rvert^2 \nonumber\\
    +&4\left(\frac{ f_{T,\phi}}{f_\phi}\right)^2\left(2\frac{m_\tau^2}{m_\phi^2}-1-\frac{m_\phi^2}{m_\tau^2}\right)\left\lvert c^{sT}_{\tau\ell X}\right\rvert^2 \bigg\},
    \label{eq:tau2phi}
\end{align}
where here and in all the above expressions $f_M (f_{T,M})$ stand for the (transverse) decay constants of the mesons, and the pseudoscalar matrix element $h^{q}_{\eta^{(')}}$ is related to the axial ($f^{q}_{\eta^{(')}}$) and gluonic ($a_{\eta^{(')}}$) elements through the Ward identity:
\begin{equation}
    h^{q}_{\eta^{(')}}=m_{\eta^{(')}}^2 f^{q}_{\eta^{(')}} + a_{\eta^{(')}}\,.
\end{equation}
Their numerical values are given in Table~\ref{tab:nuclear_matrix_elements}.

\begin{table}[t!]
\begin{center}
{\small
\renewcommand{\arraystretch}{1.3}
\begin{tabular}{|lcl|}
\hline
Decay constant & Value &
\\
\hline
$f_\pi$&130.2(1.2) MeV &\cite{Workman:2022ynf}
\\
$f_\eta^{u}/f_\pi=f_\eta^{d}/f_\pi$ & 0.77&\cite{Bali:2021qem}
\\
$f_\eta^{s}/f_\pi$ & -1.17 &\cite{Bali:2021qem}
\\
$a_\eta$ & -0.0170(10) GeV$^3$&\cite{Bali:2021qem}
\\
$f_{\eta'}^{u}/f_\pi=f_{\eta'}^{d}/f_\pi$ & 0.56&\cite{Bali:2021qem}
\\
$f_{\eta'}^{s}/f_\pi$ & 1.50 &\cite{Bali:2021qem}
\\
$a_{\eta'}$ & -0.0381(84) GeV$^3$&\cite{Bali:2021qem}
\\
$f_\omega$ & 198 MeV &\cite{Coloma:2020lgy}
\\
$f_{T,\rho}/f_\rho \sim f_{T,\omega}/f_\omega$&0.76(4)&\cite{Jansen:2009hr}
\\
$f_\phi$ & 227 MeV &\cite{Coloma:2020lgy}
\\
$f_{T,\phi}/f_\phi$&0.750(8)&\cite{RBC-UKQCD:2008mhs}
\\
\hline
Matrix element & Value & 
\\
\hline
$g_A$ & $1.27641(56)$ & \cite{Markisch:2018ndu}
\\
$g_A^{u,p}=g_A^{d,n}$ & $0.842(12)$ & \cite{PhysRevD.98.030001,HERMES:2006jyl}
\\
$g_A^{d,p}=g_A^{u,n}$ & $-0.427(13)$ & \cite{PhysRevD.98.030001,HERMES:2006jyl}
\\
$g_A^{s,p}=g_A^{s,n}$ & $-0.085(18)$ & \cite{PhysRevD.98.030001,HERMES:2006jyl}
\\
\hline
$\tilde{a}_N\simeq-m_N (g_{A}^{u,p} + g_{A}^{u,n})$ & $-0.39(12)$ GeV & \cite{Hoferichter:2022mna}
\\
\hline
$g_S^{u,p}=g_S^{d,n}$ & $0.0208(15)\frac{m_N}{m_u}$ & \cite{Hoferichter:2015dsa}
\\
$g_S^{d,p}=g_S^{d,p}$ & $0.0411(2.8)\frac{m_N}{m_d}$ & \cite{Hoferichter:2015dsa}
\\
$g_S^{s,p}=g_S^{s,n}$ & $0.043(20)\frac{m_N}{m_s}$ & \cite{Durr:2015dna,Yang:2015uis,Abdel-Rehim:2016won,Bali:2016lvx}
\\
\hline
$f_{1,T}^{u,p}=f_{1,T}^{d,n}$ & $0.784(28)$ & \cite{Gupta:2018lvp}
\\
$f_{1,T}^{d,p}=f_{1,T}^{u,n}$ & $-0.204(11)$ & \cite{Gupta:2018lvp}
\\
$f_{1,T}^{s,p}=f_{1,T}^{s,n}$ & $-0.0027(16)$ & \cite{Gupta:2018lvp}
\\
\hline
\end{tabular}

}
\end{center}
\caption{Meson decay constants and nuclear matrix elements. Due to the lack of lattice data on the transverse decay constant of the $\omega$-meson, we estimate $f_{T,\omega}/f_\omega$ to be equal to $f_{T,\rho}/f_\rho$. For the gluon element $\tilde{a}_N$ there is currently no {\it ab initio} computation, so only estimations are available. As in Ref.~\cite{Hoferichter:2022mna}, we use the FKS~\cite{Feldmann:1998vh} estimate, whose $30\%$ uncertainty is motivated by $1/N_c$ corrections. Notice also the large uncertainty in $g_S^{s,N}$ arising from a tension between phenomenological and lattice computations.}
\label{tab:nuclear_matrix_elements}
\end{table}

In analogy to the $\rho$ resonant contribution being included in the full $\tau\rightarrow \ell \pi^+ \pi^-$ decay, the $\tau\rightarrow \ell K K$ process would receive contributions from the vector resonances, particularly from the $\phi$.
While this decay could provide complementary information to the other observables listed above, as it is sensitive to all isospin combinations~\cite{Celis:2014asa,Cirigliano:2021img}, we will not include it in our analysis since the necessary form factors have large uncertainties.

\subsection[$\mu-e$ conversion in nuclei]{$\boldsymbol{\mu-e}$ conversion in nuclei}
\label{subsec:mueconv}
Operators containing a muon, an electron and two light quarks can induce $\mu-e$ conversion in nuclei, for which very strong bounds exist. In particular, this process has been searched for in four different elements (S, Ti, Pb and Au), as shown in Table~\ref{Tab:LFVlimits}.

Most studies~\cite{Raidal:1997hq,Cirigliano:2009bz,Crivellin:2017rmk,Davidson:2022nnl,Plakias:2023esq} focus on the spin-independent (SI) conversion rate since this contribution is enhanced by the coherent sum of all nucleons and thus grows with the mass number of the nucleus considered. However, this SI conversion rate only receives contributions from vector and scalar operators\footnote{As well as the dipole operator, but we will neglect its contribution given the strong bound from $\mu \to e \gamma$.} as:
\begin{equation}
    \text{CR}\left(\mu \rightarrow e\right)_{\text{SI}}=\dfrac{32 G_F^2 m_\mu^5}{\Gamma_{\text{cap}}}\sum_{X=L,R}\left\lvert V^{(p)} c^{pV}_{ X}+V^{(n)} c^{nV}_{ X}+S^{(p)} c^{pS}_{X}+S^{(n)} c^{nS}_{X}\right\rvert^2,
    \label{eq:mueconvSI}
\end{equation}
where $V^{(p),(n)}$ and $S^{(p),(n)}$ are, respectively, the vector and scalar overlap integrals for the proton and neutron, which have been evaluated in~\cite{Kitano:2002mt,Cirigliano:2009bz,Borrel:2024ylg}. 
In our analysis, we will use the latest results of Ref.~\cite{Borrel:2024ylg}.
Moreover, the proton and neutron couplings are given by:
\begin{align}
        c^{pV}_{X}&=2 c^{uV}_{\mu e X} + c^{dV}_{\mu e X}\,,
        &c^{pS}_{X}&=\sum_q\left( g_S^{q,p} c^{qS}_{\mu e X}+\frac{m_\mu}{m_N}f_{1,T}^{q,p} c^{qT}_{\mu e X}\right),\\
        c^{nV}_{X}&= c^{uV}_{\mu e X} +2 c^{dV}_{\mu e X}\,,
        &c^{nS}_{ X}&=\sum_q \left( g_S^{q,n} c^{qS}_{\mu e X}+\frac{m_\mu}{m_N}f_{1,T}^{q,n} c^{qT}_{\mu e X}\right),
\label{eq:SI_effective_coef}
\end{align}
where the finite recoil contribution from tensor operators in the scalar combinations has been included~\cite{Fitzpatrick:2012ix,Cirelli:2013ufw}. Even though this contribution is suppressed by $m_\mu/m_N$, this finite recoil term can provide useful constraints. The nuclear scalar form factors can be evaluated from pion-nucleus scattering~\cite{Crivellin:2013ipa} and from isospin breaking corrections extracted from the proton-neutron mass splitting~\cite{Brantley:2016our,Gasser:2020mzy}. Their values are given in Table~\ref{tab:nuclear_matrix_elements}.

Eq.~(\ref{eq:mueconvSI}) shows that the overlap integrals play a critical role in defining the constrained and flat directions of WCs and, as such, their uncertainties can have qualitative repercussions on a global fit. This is in stark contrast to the $\tau-\ell$ sector, in which the constrained directions are  less prone to uncertainties since they arise from criteria such as isospin or CP.
When all uncertainties are neglected, these overlap integrals define independent directions in WC space for each of the 4 nuclei for which bounds are available and thus constrain the 4 operator combinations in Eqs.~\eqref{eq:SI_effective_coef}. However, as pointed out in Ref.~\cite{Davidson:2018kud}, once the nuclear uncertainties are accounted for (at the $5-10\%$ level) some of these directions may become parallel in WC space.  
This has the devastating effect of reducing the number of constraints that are available, leading to new flat directions that were not present in the analysis without nuclear uncertainties. We will quantify this effect later when performing the global analysis of the $\mu-e$ sector.

On the other hand, axial, pseudoscalar and tensor operators can mediate spin-dependent (SD) $\mu-e$ conversion in nuclei with spin~\cite{Cirigliano:2017azj,Davidson:2017nrp,Hoferichter:2022mna}. While generally subdominant without the coherent enhancement over all nucleons, it is still useful to consider the SD contribution since the stringent experimental bounds on $\mu-e$ conversion would still translate into meaningful constraints for these kind of operators.

One of the main limitations of SD conversion versus SI conversion is that the former can only happen in nuclei with spin. Since nuclear pairing combines the nucleons in spinless pairs, nuclei with even mass number are spinless and therefore no bound from SD conversion can be derived. Moreover, the spin structure of heavy nuclei, such as Au or Pb, is challenging to quantify from the theory side. As a consequence, titanium is the only phenomenologically viable element so as to derive limits from SD conversion, having two isotopes with spin: $^{47}$Ti and $^{49}$Ti, with relative abundances of 7.44$\%$ and 5.41$\%$ respectively.

For an isotope of spin $J$ and keeping only the first multipole\footnote{Higher multipoles can be safely neglected, as their contributions are much suppressed.}, the spin-dependent conversion rate reads~\cite{Hoferichter:2022mna}:
\begin{equation}
    \text{CR}(\mu\rightarrow e)_{\rm SD}=\dfrac{32 m_\mu^5\alpha^3Z^3}{\pi \Gamma_{\rm cap}(2J+1)}\left(\dfrac{Z_{eff}}{Z}\right)^4\sum_{X=L,R}\,\sum_{\tau=\mathcal{T},\mathcal{L}}\left\lvert\sqrt{S^\tau_{00}}C^{0\tau}_X+\sqrt{S_{11}^\tau}C^{1\tau}_X\right\rvert^2,
    \label{eq:mueconvSD}
\end{equation}
where $S_{ij}^\tau$, $C^{\tau}_0$ and $C^{\tau}_1$ are, respectively, the spin structure factors, and the isoscalar and isovector combinations for the transverse ($\tau=\mathcal{T}$) and longitudinal ($\tau=\mathcal{L}$) modes. The coefficients are defined as:
\begin{align}
    C^{i\mathcal{T}}_X&=\left(1+\delta'\right)^{i} c^{iA}_X \pm 2 c^{iT}_{X}\,, \label{eq:SD_effective_coef1}\\
    C^{i\mathcal{L}}_X&=\left(1+\delta''\right)^{i} c^{iA}_X \pm 2 c^{iT}_{X}-\frac{m_\mu}{2 m_N} c^{iP}_X\,,
    \label{eq:SD_effective_coef2}
\end{align}
where the relative sign $\pm$ refers to $X=L/R$ and $\delta'$ and $\delta''$ are corrections arising from several nuclear effects~\cite{Hoferichter:2020osn}. We assume the values $\delta'=-0.28(5)$ and $\delta''=-0.44(4)$ from Ref.~\cite{Hoferichter:2022mna}.

The index $i=0,1$ refers to the isoscalar and isovector basis, which is related to the proton and neutron basis as 
\begin{equation}
    c^{0x}_X=\dfrac{c^{px}_X + c^{nx}_X}{2}\,,\hspace{1cm}c^{1x}_X=\dfrac{c^{px}_X - c^{nx}_X}{2}\,,\hspace{0.5cm}\text{for}\hspace{0.5cm}X=L,R,\hspace{0.5cm}x=A,P,T,
\end{equation}
with the proton and neutron axial, pseudoscalar and tensor couplings given by
\begin{align}
        c^{pA}_{X}&=\sum_q g_A^{q,p}c^{qA}_{\mu e X}\,,
        &c^{pP}_X&=\sum_q\frac{m_N}{m_q} g_5^{q,p}c^{qP}_{\mu e X}\,,
        &c^{pT}_X&=\sum_q f^{q,p}_{1,T}c^{qT}_{\mu e X}\,,
        \\
        c^{nA}_{X}&=\sum_q g_A^{q,n}c^{qA}_{\mu e X}\,,
        &c^{nP}_X&=\sum_q\frac{m_N}{m_q} g_5^{q,n}c^{qP}_{\mu e X}\,,
        &c^{nT}_X&=\sum_q f^{q,n}_{1,T}c^{qT}_{\mu e X}\,,
        \label{eq:nucleon_coeff}
\end{align}
in terms of the the nucleon matrix elements $\{g_A, g_5,f_{1,T}\}^{q,N}$, whose numerical values are given in Table~\ref{tab:nuclear_matrix_elements}.
The pseudoscalar matrix elements are related to the axial ones via the Ward identity:
\begin{equation}
    g_5^{q,N}=g_A^{q,N}+\dfrac{\tilde{a}_N}{2m_N}\,,
    \label{eq:pseudoscalar_axial_relation}
\end{equation}
being $\tilde{a}_N$ the gluonic matrix element. 
Finally, the spin structure factors $S_{ij}^\tau$, when considering just the first multipole, read:
\begin{align}
    \sqrt{S_{00}^{\mathcal{T}}}&=\mathcal{F}_{+}^{\Sigma'_1}(m_\mu^2)\,,
    &\sqrt{S_{00}^{\mathcal{L}}}&=\mathcal{F}_{+}^{\Sigma''_1}(m_\mu^2)\,,\\
    \sqrt{S_{11}^{\mathcal{T}}}&=\mathcal{F}_{-}^{\Sigma'_1}(m_\mu^2)\,,
    &\sqrt{S_{11}^{\mathcal{L}}}&=\mathcal{F}_{-}^{\Sigma''_1}(m_\mu^2)\,,
\end{align}
where the fit functions $\mathcal{F}^{\Sigma^{'('')}_1}_{\pm}(u)$ are given in Ref.~\cite{Hoferichter:2022mna}.

Consequently, for each lepton chirality and regardless on how many isotopes we have data on, at most 4 directions can be constrained by SD conversion, namely the 4 WC combinations in Eqs.~\eqref{eq:SD_effective_coef1}-\eqref{eq:SD_effective_coef2}: $C^{0\mathcal{T}}_X$, $C^{1\mathcal{T}}_X$, $C^{0\mathcal{L}}_X$ and $C^{1\mathcal{L}}_X$. Consequently, the parameter space of axial, pseudoscalar and tensor operators cannot be fully probed with only SD $\mu-e$ conversion and complementary probes, such as meson decays, are needed\footnote{This seems at odds with the conclusions of Ref.~\cite{Hoferichter:2022mna}. However, even though higher multipoles were considered in that work, the only combinations of operators contributing are the four in Eqs.~\eqref{eq:SD_effective_coef1}-\eqref{eq:SD_effective_coef2}. Thus, even these higher orders are not able to provide additional information.}.

The 4 independent directions to which SD constraints can be sensitive to can in principle be bounded already through $^{47}$Ti and $^{49}$Ti data. Indeed, Eq.~\eqref{eq:mueconvSD} shows that, for a fixed lepton chirality, each isotope constraints 2 directions in WC space, since each of the two modes ($\mathcal{T}$ and $\mathcal{L}$) contributes incoherently. Consequently, the two Ti isotopes are able to constrain the 4 WC combinations to which SD conversion is sensitive to. 

Summing up, $\mu-e$ conversion data would constrain on 8 different coefficient combinations per chirality, 4 from SI and 4 from SD conversion, when nuclear uncertainties are neglected. Assessing how this number is reduced in a more realistic scenario requires a careful and systematic treatment of the nuclear quantities that control the directions and their respective uncertainties. We will investigate this issue in scenarios for which bounds on all Wilson coefficients can, in principle, be derived. In particular, we will compare the results of global fits to the data when uncertainties are properly included as nuisance parameters to the naive fit without them so as to gauge their impact. 

\subsection[Meson decays: $M\rightarrow \mu e$]{Meson decays: $\boldsymbol{M\rightarrow \mu e}$}
\label{subsec:mesondecays}

Fully leptonic cLFV decays of light pseudoscalar mesons $\pi^0$, $\eta$ and $\eta'$ (see Table~\ref{Tab:LFVlimits}) provide complementary probes of $\mu-e$ cLFV operators involving quarks, as they constrain different WC combinations than $\mu-e$ conversion.
In particular, this kind of decays are sensitive to axial and pseudoscalar coefficients, accessible only via the subdominant SD contribution that, as we discussed above, does not provide enough information as to constrain them all.

Unlike the $\tau-\ell$ sector, we will not assume isospin symmetry here so as not to artificially lift flat directions, since the scalar and pseudoscalar nuclear matrix elements contributing to $\mu-e$ conversion are extracted without that assumption. The expressions for the branching ratios as a function of the WCs are the following~\cite{Hoferichter:2022mna}:
\begin{align}
    &\text{BR}\left(\pi^0 \rightarrow\mu^\mp e^\pm\right) 
    = \dfrac{G_F^2f_\pi^2m_\mu^2\big(m_\pi^2 - m_\mu^2\big)^2}{4\pi \Gamma_{\pi^0} m_{\pi}^3}
    \hspace{-1cm}&&
    \sum_{X=L,R}\Bigg|
    c^{uA}_{\mu e X}-c^{dA}_{\mu e X} 
    \mp \frac{m_\pi^2}{2m_\mu}\left(\frac{c^{uP}_{\mu e X}}{m_u }- \frac{c^{dP}_{\mu e X}}{m_d}\right)\Bigg|^2,
    \label{eq:pi02mue}\\
    &\text{BR}\left(\eta \rightarrow\mu^\mp e^\pm\right)
    = \dfrac{G_F^2f_{\pi}^2m_\mu^2\big(m_\eta^2 - m_\mu^2\big)^2}{4\pi \Gamma_{\eta} m_{\eta}^3}
    \hspace{-1cm}&&
    \sum_{X=L,R}\Bigg|\dfrac{f_\eta^{u}}{f_\pi}(c^{uA}_{\mu e X}+c^{dA}_{\mu e X})+\dfrac{f_\eta^{s}}{f_\pi} c^{sA}_{\mu e X} 
    \nonumber\\
    &&&\mp \frac{h_\eta^{u}}{2f_\pi m_\mu}
    \left(\frac{c^{uP}_{\mu e X}}{m_u} + \frac{c^{dP}_{\mu e X}}{m_d}\right)\mp \frac{h_\eta^{s}}{2f_\pi m_\mu m_s}c^{sP}_{\mu e X}\Bigg|^2,
    \label{eq:eta2mue}\\
    &\text{BR}\left(\eta' \rightarrow\mu^\mp e^\pm\right) 
    = \dfrac{G_F^2f_{\pi}^2m_\mu^2\big(m_{\eta'}^2 - m_\mu^2\big)^2}{4\pi \Gamma_{\eta'} m_{\eta'}^3}
    \hspace{-.75cm}&&
    \sum_{X=L,R}\Bigg|\dfrac{f_{\eta'}^{u}}{f_\pi}(c^{uA}_{\mu e X}+c^{dA}_{\mu e X})+\dfrac{f_{\eta'}^{s}}{f_\pi} c^{sA}_{\mu e X}
    \nonumber\\
    &&&\mp \frac{h_{\eta'}^{u}}{2f_\pi m_\mu}
    \left(\frac{c^{uP}_{\mu e X}}{m_u} + \frac{c^{dP}_{\mu e X}}{m_d}\right)\mp \frac{h_{\eta'}^{s}}{2f_\pi m_\mu m_s}c^{sP}_{\mu e X}\Bigg|^2,
    \label{eq:etaprime2mue}
\end{align}
where the matrix elements are given in Table~\ref{tab:nuclear_matrix_elements}.

In principle, bounds on these three decays would allow to constrain 6 different combinations of WCs, given the different relative sign of the pseudoscalar contribution depending on the lepton charges. In practice, however, some of the experimental limits shown in Table~\ref{Tab:LFVlimits} translate into extremely weak bounds. Even when considering only one operator at a time, the $\eta$ and $\eta'$ limits imply bounds larger than $\mathcal O(10)$ for the WCs of axial operators, since their contributions  are chirality-suppressed. For the pseudoscalar operators the situation is marginally better and the constraint on the $\eta$ decay translates into an $\mathcal O(10^{-1})$ bound, while the corresponding $\eta'$ constraint is still marginal. Thus, we will only add to the global fit the decays of the $\pi^0$ and the $\eta$.  

Lastly, notice that none of the previous observables probing cLFV in the $\mu-e$ sector are sensitive to the vector operator involving the $s$-quark. This operator could in principle be constrained by cLFV decays of vector mesons such as the $\phi$-meson:
\begin{equation}
    \text{BR}\left(\phi\to \mu^{\mp} e^{\pm}\right)=\dfrac{2 G_F^2 f_\phi^2 m_\phi^3}{3\pi \Gamma_\phi}\sum_{X=L,R}\left\{\left\lvert c^{sV}_{\mu e X}\right\rvert^2+2\left(\frac{f_{T,\phi}}{f_\phi}\right)^2\left\lvert c^{sT}_{\mu e X}\right\rvert^2\right\}.
\end{equation}
Unfortunately, the available constraint on this decay~\cite{Achasov:2009en} leads to a rather weak bound $\mathcal O(10^2)$ on the corresponding WCs, thus leaving $c^{sV}_{\mu e X}$ essentially unconstrained. Consequently, we will not consider this bound for the purposes of our global fit.

\section{LEFT analysis}
\label{sec:LEFT}

\begin{figure}[t!]
    \centering
    \includegraphics[width=0.95\textwidth]{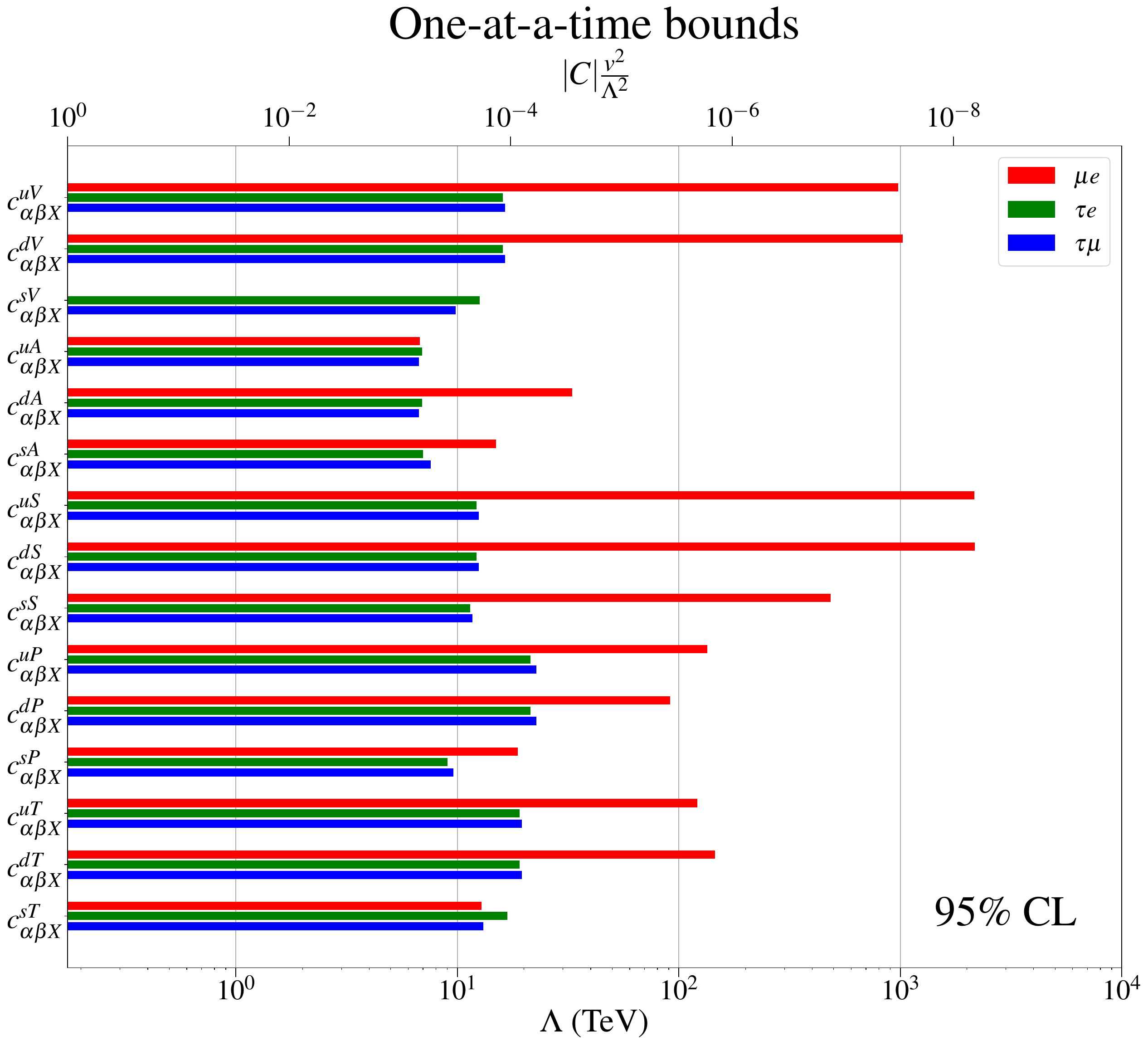}
    \caption{Current $95\%$ CL bounds on LEFT cLFV operators with quarks considering only one operator at a time (see Fig.~\ref{fig:circular_barplot} for the rest of operators). Missing bars indicate that there is no (relevant) bound for those operators at present. }
    \label{fig:one-at-a-time_bounds}
\end{figure}

From the expressions for the cLFV observables shown in the previous section, we first consider only one of the LEFT operators in Table~\ref{tab:LEFT} at a time and derive upper limits on their WCs (or equivalently lower limits for the scale $\Lambda$ they probe). 
We present these bounds in Fig.~\ref{fig:one-at-a-time_bounds}. 

As expected, all the constraints in the $\tau-\mu$ and $\tau-e$ sectors are similarly strong, around $10^{-3}$ and $10^{-4}$ for the WCs, since the constraints from BaBar and Belle on the different cLFV decays of the $\tau$ are all of the same order (see Table~\ref{tab:LEFT}). Notably, the bounds on the pseudoscalar operators are slightly stronger since they circumvent the chirality flip suppression otherwise required for the decays into pseudoscalar mesons. In the $\mu-e$ sector there is a much larger disparity on the order of magnitude of the constraints depending on which is the observable that dominates the bound. Indeed, the strongest constraints between $10^{-7}$ and $10^{-8}$ correspond to the scalar and vector couplings since these induce the, resonantly enhanced, SI contribution to $\mu-e$ conversion. The tensor structures are also strongly bounded at the level of $10^{-5}$ to $10^{-6}$ since they may also mediate the SI conversion via their finite recoil contribution with an additional $m_\mu/m_N$ suppression. Bounds on axial and pseudoscalar operators rather stem from their contribution to the SD conversion, which lacks the coherent enhancement. Nevertheless, pseudoscalar operators are also bounded at the $10^{-5}$ level given the enhancement with $m_N/m_q$ in Eq.~\eqref{eq:nucleon_coeff}, while the bounds on the axial structure is order $10^{-4}$. For all Lorentz structures, the constraints on operators involving the $s$ quark are between one and two orders of magnitude weaker given their suppressed nucleon matrix elements with no bound at all for the vector operator. This justifies the simplified scenario without $s$ quarks studied in Section~\ref{sec:SMEFT-s}, since flat directions involving $s$ quarks would need to overcome this matrix element suppression. 

In this section we go beyond this one-at-a-time approach and attempt to derive global bounds on these LEFT operators discussing, both qualitatively and quantitatively, how the results change when all operators are considered at the same time. For this, and also for the results presented in the next sections, we build the $\chi^2$ function adding the constraints from all the observables discussed in section~\ref{sec:obs}, where we have assumed the best fits of all upper bounds to be at zero and a Gaussian distribution in all cases. Since the parameter spaces describing the different EFTs we will confront with the observables are rather sizable, we will explore them via MCMC sampling. Furthermore, given the existence of both flat directions and of bounds on different operator combinations of very disparate orders of magnitude, we adjust the proposal function of the MCMC sampling to the directions along which we expect the most stringent bounds, based on the different observables, and with matching steps. Otherwise, it would be very easy to miss the extremely thin directions that are poorly bounded or even unconstrained and conclude that stronger bounds applied for all operators. We believe this might explain some differences between our results and others present in the literature. We then derive frequentist confidence intervals for each WC profiling over all others from the $\chi^2$ values obtained after the MCMC exploration of the parameter space. This will be summarized as $95\%$ C.L. bands in the different figures shown throughout the paper. We will also show the different 2-parameter depictions of the frequentist confidence intervals in triangle plots in Appendix~\ref{App:results} to better showcase the rather strong degeneracies as well as the impact of nuclear uncertainties in $\mu-e$ conversion constraints. Finally, we also provide approximate correlation matrices so as to be able to implement our results taking into account properly the very strong correlations found in our analysis.

Finally, notice that all of the processes discussed in the previous section constrain simultaneously operators having both chiralities for the lighter lepton involved, since their eventual interference is chirally suppressed.
Therefore, in the following we will present our discussion for a fixed chirality, keeping in mind that the qualitative arguments apply to operators of both light lepton chiralities.

\subsection[$\tau-\ell$ sector]{$\boldsymbol{\tau-\ell}$ sector}

\begin{table}[t!]
\centering
    \begin{tabular}{|c|c|c|c|c|c|c|c|c|c|c|c|c|c|c|c|}
    \hline 
    &\multicolumn{5}{c|}{}&\multicolumn{5}{c|}{}&\multicolumn{5}{c|}{}\\[-2.ex]
    \multirow{2}{*}{Observable}&\multicolumn{5}{c|}{$c_{\tau\ell}^{ux}-c_{\tau\ell}^{dx}$} 
    &\multicolumn{5}{c|}{$c_{\tau\ell}^{ux}+c_{\tau\ell}^{dx}$}&\multicolumn{5}{c|}{$c_{\tau\ell}^{sx}$}\\[1ex]
    \cline{2-16} &&&&&&&&&&&&&&&\\[-2.5ex]
    &$V$&$A$&$S$&$P$&$T$   &   $V$&$A$&$S$&$P$&$T$   &   $V$&$A$&$S$&$P$&$T$\\
    \hline
    $\tau\rightarrow\ell \pi^0$&&1\cellcolor{CornflowerBlue}&&1\cellcolor{CornflowerBlue}&
    &&&&& 
    &&&&&\\
    \hline
    $\tau\rightarrow\ell \eta$&&&&&
    &&2\cellcolor{Red}&&2\cellcolor{Red}&
    &&2\cellcolor{Red}&&2\cellcolor{Red}&\\
    \hline
    $\tau\rightarrow\ell \eta'$&&&&&
    &&3\cellcolor{Yellow}&&3\cellcolor{Yellow}&
    &&3\cellcolor{Yellow}&&3\cellcolor{Yellow}&\\
    \hline
    $\tau\rightarrow\ell \omega$&&&&&
    &4\cellcolor{Tan}&&&&5\cellcolor{LimeGreen} 
    &&&&&\\\hline
    $\tau\rightarrow \ell\pi^+\pi^-$&6\cellcolor{Gray}&&&&7\cellcolor{YellowOrange}&&&8\cellcolor{Lavender}&&&&&9\cellcolor{Mulberry}&&\\\hline
    $\tau\rightarrow\ell \phi$&&&&&
    &&&&& 
    &10\cellcolor{NavyBlue}&&&&11\cellcolor{Green}\\
    \hline
    \end{tabular}
\caption{Summary of combinations of LEFT WCs constrained by semileptonic $\tau$ decays, given in the isospin basis $u\pm d$. 
Filled boxes indicate to which WCs each observable is sensitive to, with same color and number indicating coherent contributions between those WCs. 
Thus, different colors/numbers correspond to the 11 independent constraints in the $\tau-\ell$ sector.}
\label{tab:tauLFV_constrained_directions}
\end{table}

As shown in Table~\ref{tab:LEFT}, and for a fixed chirality of the lighter lepton $\ell$, there are 5 operators per quark flavour, each with a different Lorentz structure (V, A, S, P, T).
In our three flavour approach with operators only involving $u$, $d$ and $s$, this means a total of 15 indepenndent operators.

As discussed in section~\ref{subsec:tauLFVobservables}, these operators are probed by several cLFV semileptonic $\tau$ decays. Each decay to a pseudoscalar meson ($\pi^0$, $\eta$, $\eta'$) probes a different combination of axial and pseudoscalar WCs -- see Eqs.~\eqref{eq:tau2pi0}, \eqref{eq:tau2eta} and \eqref{eq:tau2etaprime} --; the decay $\tau\rightarrow\ell \pi^+\pi^-$ independently constrains four different WC combinations -- Eq.~\eqref{eq:tau2pipi} --; and finally the $\tau\rightarrow\ell\omega$ and $\tau\rightarrow\ell\phi$ decays provide four additional constraints -- Eqs.~\eqref{eq:tau2omega} and \eqref{eq:tau2phi}--.
The counting of these constrained directions can be visualized in Table~\ref{tab:tauLFV_constrained_directions}.

All in all, the $\tau$ cLFV semileptonic decays under consideration probe 11 different combinations of WCs.
Since the LEFT contains 15 operators that can contribute to these processes, there are 4 flat directions (F1, F2, F3, F4) that remain unconstrained:
\begin{itemize}
    \item[$\bullet$] F1: $c^{uS}_{\tau\ell X}-c^{dS}_{\tau\ell X}$.
    \item[$\bullet$] F2: one combination of $c^{uA}_{\tau\ell X}-c^{dA}_{\tau\ell X}$ and $c^{uP}_{\tau\ell X}-c^{dP}_{\tau\ell X}$.
    \item[$\bullet$] F3, F4: two combinations of $\big\{c^{qA}_{\tau\ell X}\big\}_{q=u+d,s}$ and $\big\{c^{qP}_{\tau\ell X}\big\}_{q=u+d,s}$.
\end{itemize}

The direction F1 can be easily understood since the only constraints on scalar WCs come from $\tau\rightarrow \ell\pi^+\pi^-$, which only constrains the isoscalar combination and $c^{sS}_{\tau\ell X}$, leaving the isovector combination unconstrained. The direction F2 arises from the fact that $\tau\to\ell\pi^0$ is the only probe of isovector axial and pseudoscalar operators, therefore the combination that cancels Eq.~\eqref{eq:tau2pi0} is left unconstrained. Analogously, F3 and F4 are a consequence that isoscalar and $s$-quark axial and pseudoscalar operators are only constrained by $\tau\to\ell\eta$ and $\tau\to\ell\eta'$, thus leaving two unbounded directions along which both Eqs.~\eqref{eq:tau2eta} and~\eqref{eq:tau2etaprime} vanish at the same time.

\begin{figure}[t!]
    \centering
    \includegraphics[width=0.8\textwidth]{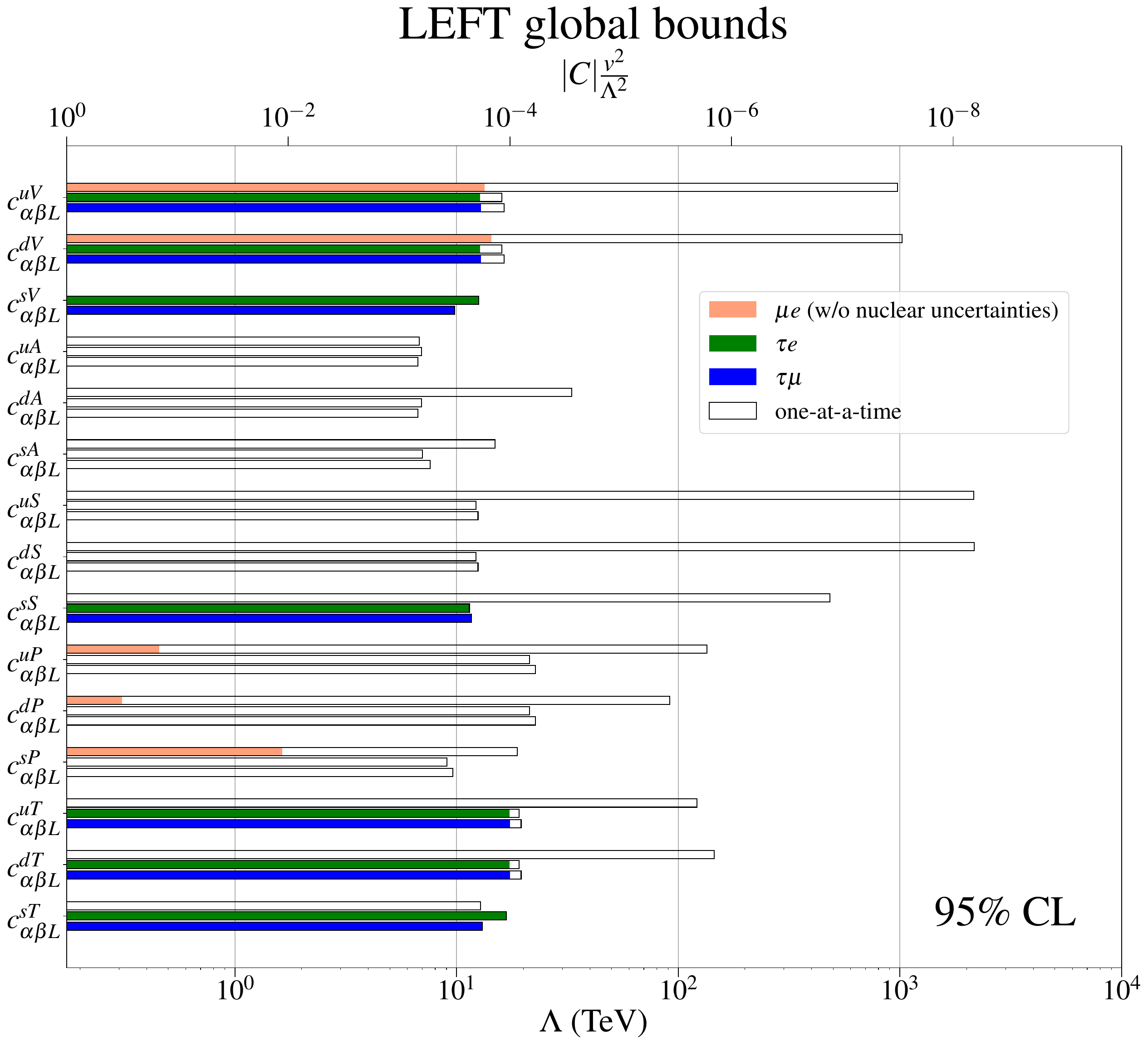}
    \caption{Current 95\%~CL global bounds on LEFT cLFV operators involving the three lightest quarks. All operators are considered at the same time and their WCs are profiled over to obtain individual bounds. 
    Missing bars indicate that there is no (relevant) global bound for those WCs.
    For easier comparison, we depict as empty bars the one-at-a-time constraints of Fig.~\ref{fig:one-at-a-time_bounds}.}
    \label{fig:LEFT_global_bounds}
\end{figure}

Consequently, a global analysis is not able to simultaneously constrain all of the coefficients, since combinations along the flat directions are unconstrained. Nevertheless, the flat directions presented above are restricted to axial, scalar and pseudoscalar WCs. Thus, vector and tensor coefficients can all be unambiguously constrained through a global fit in spite of the flat directions. 
In particular, $\tau\rightarrow\ell\pi^+\pi^-$ and $\tau\rightarrow\ell\omega$ respectively constrain the isovector and isoscalar combinations of vector and tensor WCs, while $\tau\rightarrow\ell\phi$ independently constrains the $s$-quark vector and tensor. Additionally $\tau\rightarrow\ell\pi^+\pi^-$ also independently constraints the scalar $s$-quark operator. Summing up, a global analysis can constrain 7 out of the 15 LEFT operators  involving flavour change in the $\tau$ sector. These constraints, result of the global fit, are shown in Fig.~\ref{fig:LEFT_global_bounds}.

Finally, it should be noted that adding $\tau\rightarrow \ell K^+K^-$ would help closing the F1 direction, however we did not include it since there are large uncertainties on the relevant form factor\footnote{For an estimation within the resonant chiral perturbation theory framework see Ref.~\cite{Husek:2020fru}.}. 
On the other hand, the flat directions involving axial and pseudoscalar operators are much more difficult to lift. This is because these operators are only involved in the decays with pseudoscalar mesons, and only the decays to $\pi^0$, $\eta$ and $\eta'$ are available, providing constraints on just 3 independent combinations.

\subsection[$\mu-e$ sector]{$\boldsymbol{\mu-e}$ sector}

In complete analogy to the $\tau-\ell$ sector, there are $15$ independent LEFT operators involving $\mu$, $e$ and two light quarks. In an idealized scenario, {\it i.e.}~neglecting nuclear uncertainties, $\mu-e$ conversion data can probe $8$ different combinations of WCs, as outlined in section~\ref{subsec:mueconv}, 4 from SI and 4 from SD $\mu-e$ conversion. 
Additionally, the $\pi^0$ and $\eta$ decays discussed in section~\ref{subsec:mesondecays} can provide (weaker) bounds on 2 additional directions\footnote{The decays to $\mu^-e^+$ and $\mu^+e^-$ would in principle lead to 4 additional constraints. However, the only meaningful constraint derived from the $\eta$ decay is on the pseudoscalar combination of operators, as the bound on the axial contribution is too weak given the chirality suppression and one of the bounds coming from the $\pi^0$ decay is redundant with the constraints from SD $\mu-e$ conversion.}. Thus, $15-10=5$ flat directions are present in this scenario. Nevertheless, as for the $\tau$ case, some Lorentz combinations can be fully constrained even in presence of these flat directions. 

Indeed, SI $\mu-e$ conversion constrains the scalar and vector couplings of the proton and neutron. For the vector operators this translates directly into the bounds on the couplings to the $u$ and $d$ quarks with the same vector structure shown in Fig.~\ref{fig:LEFT_global_bounds}. However, there is a flat direction (F1) corresponding to the $s$-quark vector operator, as none of the observables analyzed in the global fit has any sensitivity to it. In particular, in the SI conversion rate (which is the only observable sensitive to vector operators), the nucleon vector form factor for the $s$-quark vanishes due to the vector current conservation.

On the other hand, the scalar operators of proton and neutron also constrained from SI $\mu-e$ conversion receive contributions from all three scalar operators as well as from the tensor structures through their (suppressed) contribution from the finite recoil term (see Eq.~\eqref{eq:SI_effective_coef}). As such, these two constraints are not enough to derive bounds on the scalar or tensor operators. 

Regarding the pseudoscalar operators, 2 combinations are constrained from the decays of the $\pi^0$ and $\eta \to \mu e$. Furthermore, the isoscalar combination of pseudoscalar operators can also be bounded from the isoscalar contributions to SD $\mu-e$ conversion (see Eqs.~\eqref{eq:SD_effective_coef1}-\eqref{eq:SD_effective_coef2}). Thus, global bounds can be found for all 3 pseudoscalar operators. Conversely, the 9 operators involving scalar, axial and tensor structures are bounded by the 2 constrains from the scalar couplings to the proton and neutron from SI $\mu-e$ conversion mentioned above and the 3 remaining independent combinations in Eqs.~\eqref{eq:SD_effective_coef1}-\eqref{eq:SD_effective_coef2} from SD $\mu-e$ conversion. All in all, 5 constraints for 9 operators, leaving another 4 flat directions unconstrained.  Thus, the LEFT $\mu-e$ sector has 5 flat directions: 
\begin{itemize}
    \item[$\bullet$] F1: $c^{sV}_{\mu e X}$.
    \item[$\bullet$] F2, F3, F4, F5: four combinations of $\big\{c^{qx}_{\mu e X}\big\}_{q=u,d,s}$ with $x=A,S,T$.
\end{itemize}

Summing up, through a global fit and even in absence of nuclear uncertainties, it is only possible to simultaneously constrain the pseudoscalar operators as well as the $u$ and $d$ vector structures.
This is depicted in Fig.~\ref{fig:LEFT_global_bounds}, showing the dramatic effect of performing a global analysis rather than considering just a single operator at a time. It is also interesting to see how the constraints that do survive in this global fit compare to the ones derived through the one-operator-at-a-time approach. While the bounds from the $\tau-\mu$ and $\tau-e$ sectors are only slightly changed, if at all, those on the $\mu-e$ sector are relaxed by about 4 orders of magnitude. For the pseudoscalar operators the three necessary constraints come from SD $\mu-e$ conversion, $\pi$ and $\eta \to \mu e$ decays and, as such, the profiled bound is dominated by the weakest of three, namely $\eta \to \mu e$. Conversely, in the one-operator-at-a-time approach the bounds are dominated by the most stringent observable, in this case those from $\mu-e$ conversion, explaining the huge relaxation seen in Fig.~\ref{fig:LEFT_global_bounds}. 

For the vector operators, this is instead a consequence of the different operator directions being probed by the SI transition of the different nuclei characterized by their overlap integrals in Eq.~\eqref{eq:SI_effective_coef} being very close to each other. The situation becomes even worse when uncertainties on their values are taken into account through the appropriate nuisance parameters. 
The role of the uncertainties will be fully taken into account and discussed more in detail in the following sections with more constrained scenarios where their impact is more easily investigated.

\section{SMEFT analysis}
\label{sec:SMEFT}

As we have shown in the previous section, present constraints are not enough to derive bounds on all $d=6$ LEFT operators given the several flat directions that remain unconstrained. Thus, it is interesting to consider how the situation changes when the LEFT is obtained as the low-energy description of $d=6$ SMEFT operators. As discussed in section~\ref{sec:EFT}, the $d=6$ SMEFT matching generates non-trivial correlations for the (pseudo)scalar and tensor WCs (see Eqs.~\eqref{eq:smefttoleftfisrt}-\eqref{eq:smefttoleft}):
\begin{align}
        c^{uP}_{\alpha\beta X} + c^{dP}_{\alpha\beta X} &= \mp\Big(c^{uS}_{\alpha\beta X} - c^{dS}_{\alpha\beta X}\Big)\,,
        \label{eq:SMEFTcorrelationsU}\\
        c^{uP}_{\alpha\beta X} - c^{dP}_{\alpha\beta X} &= \mp\Big(c^{uS}_{\alpha\beta X} + c^{dS}_{\alpha\beta X}\Big)\,,
        \label{eq:SMEFTcorrelationsD}\\
        c^{sP}_{\alpha\beta X}&=\pm c^{sS}_{\alpha\beta X}\,,\label{eq:SMEFTcorrelationsS}\\
        c^{dT}_{\alpha\beta X}&=c^{sT}_{\alpha\beta X}=0\,,\label{eq:SMEFTcorrelationsT}
\end{align}
where the upper (lower) sign refers to $X=L (R)$ operators. These relations reduce the number of independent operators, making them easier to constrain.

\subsection[$\tau-\ell$ sector]{$\boldsymbol{\tau-\ell}$ sector}
\label{sec:SMEFT-tau}

\begin{table}[t!]
\centering
    \begin{tabular}{|c|c|c|c|c c|c|c|c|c|c|c|}
    \hline 
    &\multicolumn{4}{c|}{}&\multicolumn{4}{c|}{}&\multicolumn{3}{c|}{}\\[-2.ex]
    \multirow{2}{*}{Observable}&\multicolumn{4}{c|}{$c_{\tau\ell}^{ux}-c_{\tau\ell}^{dx}$} 
    &\multicolumn{4}{c|}{$c_{\tau\ell}^{ux}+c_{\tau\ell}^{dx}$}&\multicolumn{3}{c|}{$c_{\tau\ell}^{sx}$}\\[1ex]
    \cline{2-12} &&&&&&&&&&&\\[-2.5ex]
    &$V$&$A$&$S$&\multicolumn{2}{c|}{$T$} &     $V$&$A$&$S$&  $V$&$A$&$S$\\
    \hline
    $\tau\rightarrow\ell \pi^0$&&1\cellcolor{CornflowerBlue}&&
    &&&&1\cellcolor{CornflowerBlue} 
    &&&\\
    \hline
    $\tau\rightarrow\ell \eta$&&&2\cellcolor{Red}&&
    &&2\cellcolor{Red}&&
    &2\cellcolor{Red}&2\cellcolor{Red}\\
    \hline
    $\tau\rightarrow\ell \eta'$&&&3\cellcolor{Yellow}&&
    &&3\cellcolor{Yellow}&&
    &3\cellcolor{Yellow}&3\cellcolor{Yellow}\\
    \hline
    $\tau\rightarrow\ell \omega$&&&&
    &5\cellcolor{LimeGreen}&4\cellcolor{Tan}&&
    &&&\\\hline
    $\tau\rightarrow \ell\pi^+\pi^-$&6\cellcolor{Gray}&&&7\cellcolor{YellowOrange}&
&&&8\cellcolor{Lavender}
&&&9\cellcolor{Mulberry}\\\hline
    $\tau\rightarrow\ell \phi$&&&&
    &&&&& 
    10\cellcolor{NavyBlue}&&\\
    \hline
    
    \hline
    \end{tabular}
\caption{Same as Table~\ref{tab:tauLFV_constrained_directions} but for when the LEFT operators are induced by low energy $d=6$ SMEFT, which has less independent WCs (see text for details). In particular, the isoscalar and isovector tensor operator map to the same SMEFT contribution, constrained by both $\tau\to\ell\omega$ and $\tau\to\ell\pi^+\pi^-$. }
\label{tab:tauLFV_constrained_directions_SMEFT}
\end{table}

Upon matching with $d=6$ SMEFT, the tensor coefficients for down-type quarks vanish, while the scalar and pseudoscalar coefficients are identical up to a sign.
This reduces the number of independent WCs from the 15 in LEFT to just 10 coefficients. On the other hand, these same correlations also reduce the number of independent constraints as some of them become redundant. In particular, since there is no tensor coefficient $c^{dT}_{\tau\ell L}$, both decays $\tau\rightarrow \ell\pi^+\pi^-$ and $\tau\rightarrow \ell\omega$ overconstrain the only tensor coefficient $c^{uT}_{\tau\ell L}$. Analogously, the coefficient $c^{sT}_{\tau\ell L}$ vanishes, rendering its bound from $\tau\rightarrow \ell\phi$ irrelevant. We display this new counting for the SMEFT schematically in Table~\ref{tab:tauLFV_constrained_directions_SMEFT}. 
All in all, in the low-energy $d=6$ SMEFT, 2 of the 11 constraints are redundant, reducing the number of independent constraints to 9. 

Compared to the LEFT scenario, Eq.~\eqref{eq:SMEFTcorrelationsS} implies that the constraint on $c^{sS}_{\tau\ell X}$ from $\tau\rightarrow \ell\pi^+\pi^-$ translates to a bound on $c^{sP}_{\tau\ell X}$ as well. Similarly, the bound on the isoscalar combination $c^{uS}_{\tau\ell X}+c^{dS}_{\tau\ell X}$ translates to a bound on the isovector combination $c^{uP}_{\tau\ell X}-c^{dP}_{\tau\ell X}$. The isovector combination of axial WCs is then constrained through the bound on $\tau\to \ell\pi^0$. This leaves unconstrained the isoscalar and $s$-quark axial coefficients, as well as the isovector combination $c^{uS}_{\tau\ell X}- c^{dS}_{\tau\ell X}$, for which there are only two remaining constraints coming from decays to the $\eta$ and $\eta'$ mesons, so only one linear combination of these three operators remains unconstrained. 
Thus, in this context, the single remaining flat direction will be:
\begin{itemize}
    \item[$\bullet$] F1: a combination of $c^{uA}_{\tau\ell X}$+ $c^{dA}_{\tau\ell X}$, $c^{sA}_{\tau\ell X}$ and $c^{uS}_{\tau\ell X}- c^{dS}_{\tau\ell X}$
\end{itemize}

\begin{figure}[t!]
    \centering
    \includegraphics[width=0.8\textwidth]{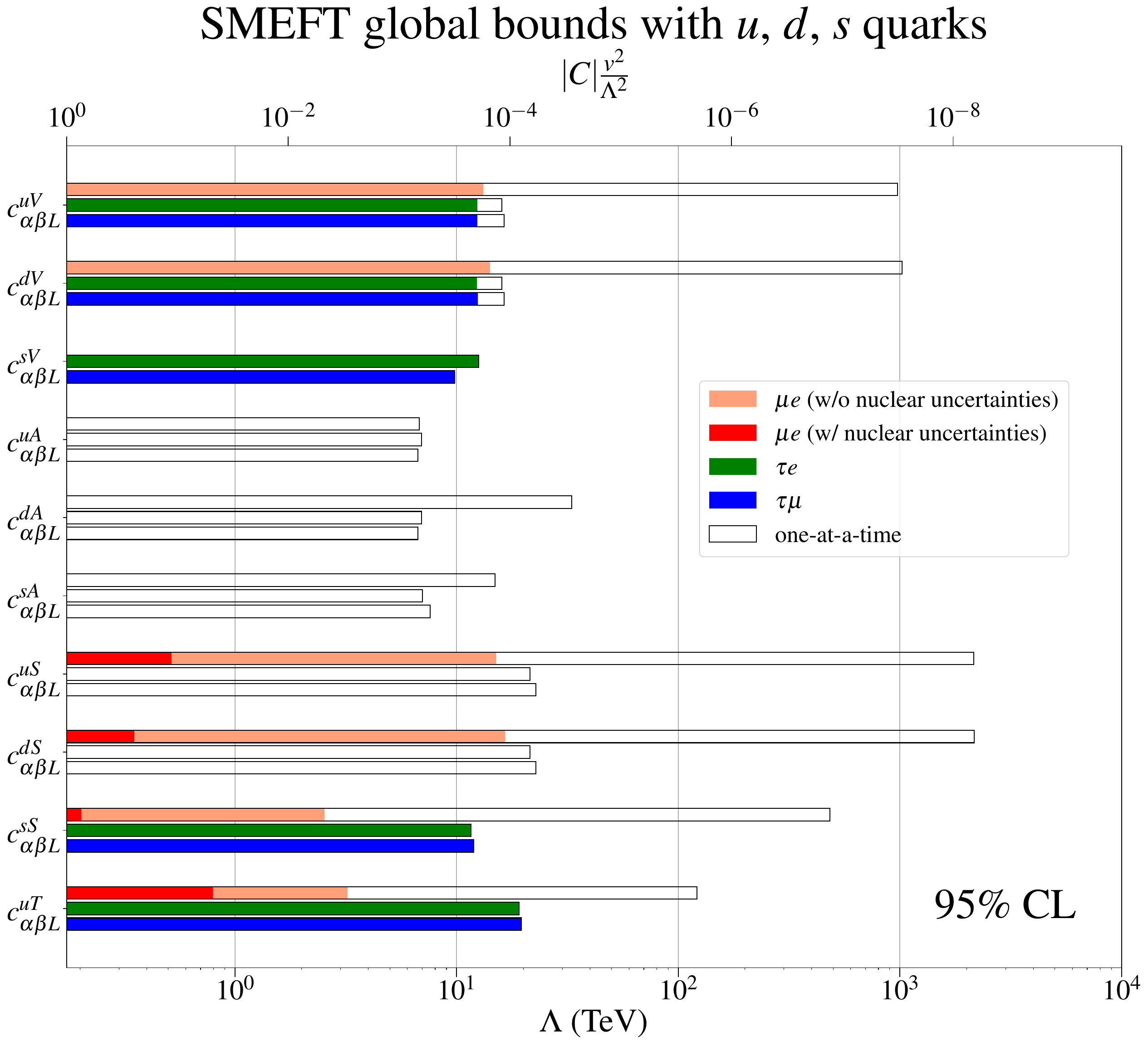}
    \caption{Current 95\% CL global bounds on the cLFV LEFT operators with the three lightest quarks induced by $d=6$ SMEFT at low energies. Color code as in Fig.~\ref{fig:LEFT_global_bounds}, but now darker red bars show the effects of including nuclear uncertainties in the $\mu-e$ analysis.
    Bounds on pseudoscalar operators are equal to the scalar ones due to the correlations in Eqs.~\eqref{eq:SMEFTcorrelationsU}-\eqref{eq:SMEFTcorrelationsS}. These global bounds, along with their correlations, are collected in appendix~\ref{App:results}.}
    \label{fig:SMEFT_global_uds_bounds}
\end{figure}

The bounds resulting from the global analysis to this framework are shown in Fig.~\ref{fig:SMEFT_global_uds_bounds}. The results seem very similar to those obtained in Fig.~\ref{fig:LEFT_global_bounds}. Indeed, the bounds on the WCs that were bounded in Fig.~\ref{fig:LEFT_global_bounds} are essentially the same in Fig.~\ref{fig:SMEFT_global_uds_bounds} and no new bars for other WCs appear. This is because, even if the correlations among WCs implied by the $d=6$ SMEFT allow to lift 3 out of the 4 flat directions present in the general LEFT scenario, the remaining flat direction involves all the WCs that were not previously bounded. Thus, even though the parameter space is in general much more strongly constraint and 3 flat directions have been lifted, when profiling over all WCs no individual bound is found for the WCs involved in the remaining flat direction. Thus, a better perspective on how constrained the parameter space is in this scenario is provided by Fig.~\ref{fig:corr_taue} in appendix~\ref{App:results} through the triangle plot with all different projections for constrained directions in the parameter space.

Finally, we note again that this flat direction could in principle be lifted by including the $\tau\rightarrow \ell K^+ K^-$ channel which, contrary to $\tau\rightarrow \ell \pi^+\pi^-$, is sensitive to all combinations of scalar operators ($\pi^+\pi^-$ is not sensitive to the isovector combination), and thus allows for a constraint on all scalar WCs. 
This would help closing the whole SMEFT parameter space in the $\tau-\ell$ sector. 
However, we do not include $\tau\rightarrow \ell K^+ K^-$ since the scalar hadronic form factors suffer from large theoretical uncertainties, as previously discussed.

\subsection[$\mu-e$ sector]{$\boldsymbol{\mu-e}$ sector}
\label{sec:SMEFT-s-mue}

Analogously to the previous section, the number of independent operators inducing cLFV in the $\mu-e$ sector with light quarks is 10 when matching with $d=6$ SMEFT. 

We discuss first the simplified (and overly optimistic) scenario of neglecting the impact of the nuclear uncertainties. As discussed before, $\mu-e$ conversion data can at most constrain $4+4=8$ combinations between SI and SD, and thus is not enough to cover the whole parameter space. 
Therefore, it is necessary to consider also meson decay bounds that, although substantially weaker, probe complementary combinations of WCs. 
This adds to a total of $4+4+2=10$ constrained directions, which, at face value, seem enough to constrain all of the 10 WCs.
However, as mentioned in the previous section, none of these observables receive contributions from $s$-quark vector currents, so the corresponding WC ($c^{sV}_{\mu e X}$) remains unbounded, and only 9 Wilson coefficients can be constrained at most. 

On the other hand, $\pi^0\rightarrow\mu e$ is sensitive to the isovector combination of axial coefficients, which is already bounded more strongly by SD $\mu-e$ conversion. Indeed, an analysis of the correlation matrix of the $8$ WC combinations probed by $\mu-e$ conversion reveals that the flat unconstrained direction corresponds to a combination of isoscalar and $s$-quark axial WCs. This direction is obviously orthogonal to the isovector combination probed by $\pi^0\rightarrow\mu e$ and thus cannot be lifted by this limit.
This means that the constraints derived for axial operators will be dominated by the $\eta\rightarrow \mu e$ bound, which, for axial operators, is very weak due to the chirality suppression.
Consequently, even though it is technically possible to simultaneously constrain all these 9 WCs, the constraints derived for axial operators are very weak, $\sim O(10)$, and do not show up among the other light red bands for the range shown in Fig.~\ref{fig:SMEFT_global_uds_bounds}. On the other hand, it is still possible to obtain meaningful global bounds for the remaining Lorentz structures, even if they are substantially weaker than those derived when considering one operator at a time.

We will now address the effects of adding nuclear uncertainties to the analysis.  Accounting for the impact of all uncertainties entering the $\mu \to e$ transition estimation requires treating as nuisance parameters many different quantities and the analysis can easily get out of hand and become numerically unfeasible. For this reason, we will implement uncertainties only in a selection of nuclear quantities that can be particularly harmful, since they may effectively reduce the number of independent constraints. These are:
\begin{itemize}
    \item[$\bullet$] Nuclear overlap integrals: these define the directions probed by SI $\mu-e$ conversion. The main source of uncertainty comes from the fact that lepton-nucleon interactions are computed at LO in $\chi$PT, see Refs.~\cite{Davidson:2017nrp,Bartolotta:2017mff,Hoferichter:2015dsa} for discussions about the size of the possible NLO corrections. We will consider $5\%$ and $10\%$ uncertainties for the overlap integrals of light and heavy nuclei, respectively. In particular, we will consider the parameters $V^{(p),(n)}$ and $S^{(p),(n)}$ in Eq.~\eqref{eq:mueconvSI} as free and independent with Gaussian priors centered around their nominal value and a $5\%$/$10\%$ uncertainty. With this parametrization, there will be a confidence level at which the freedom allowed for the overlap integrals makes two of them parallel~\cite{Davidson:2018kud} and, therefore, redundant instead of complementary at this C.L. This has a dramatic impact in the analysis, since a new flat direction appears, loosening many constraints and changing significantly the correlations among the remaining ones, as depicted in Appendix~\ref{App:results}.
    
    \item[$\bullet$] Nuclear corrections $\delta'$ and $\delta''$ to the axial contribution of SD $\mu-e$ conversion: these have relatively large uncertainties and, when they become equal, a new flat direction arises. This is due to the fact that, in this limit, both longitudinal and transverse modes in Eqs.~\eqref{eq:SD_effective_coef1}-\eqref{eq:SD_effective_coef2} would become sensitive to the exact same combination of isovector axial and tensor coefficients\footnote{Notice that the pseudoscalar operators are independently bounded from meson decays and the isoscalar contribution of SD $\mu-e$ conversion.}.
    
    \item[$\bullet$] Gluonic matrix element $\tilde{a}_N$: the naive estimation of this parameter, which would suffer from a $\sim 30\%$ uncertainty~\cite{Hoferichter:2022mna}, leads to a cancellation of the isoscalar pseudoscalar contribution to the SD transition via the relation between the pseudoscalar and axial matrix elements in Eq.~\eqref{eq:pseudoscalar_axial_relation}. Thus, taking into account this uncertainty has a sizable impact in the final results of the global fit.
\end{itemize}

The results of the global fit after including all these nuclear uncertainties are displayed in Fig.~\ref{fig:SMEFT_global_uds_bounds} with dark red bars. The comparison with the light red bars in absence of uncertainties is remarkable. The greatest impact is seen in the vector and scalar structures whose bounds mainly came from the SI contribution. Indeed, for a sufficiently high confidence level (somewhat below $95\%$ where the bounds are depicted), the directions corresponding to SI $\mu-e$ conversion in Pb and S become parallel to those of Au and Ti, respectively. This effectively means that two constrained directions are lost in the fit, and thus $\mu-e$ conversion can only constrain $6$ combinations out of the total $8$ directions probed without nuclear uncertainties. These two lost directions must be then supplemented by the two directions probed by meson decays, which entail weaker constraints. 

This is best shown by Fig.~\ref{fig:corr_mue_uds} in appendix~\ref{App:results} with two very striking features. The first is the rather strong correlations between most observables, which reduces the allowed regions to thin lines in the parameter space. This is the consequence of the extremely different orders of magnitude of the bounds that each operator can set, with more than 8 orders of magnitude difference between those coming from SI $\mu-e$ conversion to those from the $\eta \to \mu e$ decay. The second remarkable feature in Fig.~\ref{fig:corr_mue_uds} is the abrupt change from the 1 to the 2 $\sigma$ regions for several parameters. Some correlations are lost and the corresponding allowed regions become much larger when going from 1 to 2 $\sigma$ C.L. In some parameters there is also a dramatic change in the overall constrain as shown by the abrupt jump of the profiled $\chi^2$ depicted in the diagonal panels. This jump and the dramatic change in the correlations, happen close to the 95\% C.L., which is when the nuisance parameters can vary enough so as to make the directions probed by Pb and S as well as Au and Ti parallel, as discussed above.

\vspace{.5cm}

\section{SMEFT with only first generation quarks}
\label{sec:SMEFT-s}

Given that it is not possible to constrain all WCs simultaneously even when considering only operators that match to SMEFT at $d=6$, we consider the further simplification of operators involving only the first quark generation. This is motivated by the observables involved when constraining the $\mu \to e$ transitions, where the dominant contributions are always from operators with $u$ and $d$ quarks. Thus, flat directions involving $s$ quarks may be regarded as particularly fine-tuned, since they generally need to overcome this additional suppression. This is not the case for $\tau \to e$ and $\tau \to \mu$ transitions, since the $s$ content of several mesons involved in relevant $\tau$ decays is significant. Nevertheless, we also show results for the $\tau-\ell$ sector in this simplified scenario for completeness\footnote{Notice that a similar reduction of the number of free parameters would be achieved when assuming the same coupling for the $d$ and $s$ quarks, leading to qualitatively similar results.}.  

\subsection[$\tau-\ell$ sector]{$\boldsymbol{\tau-\ell}$ sector}

\begin{figure}[t!]
    \centering
    \includegraphics[width=0.8\textwidth]{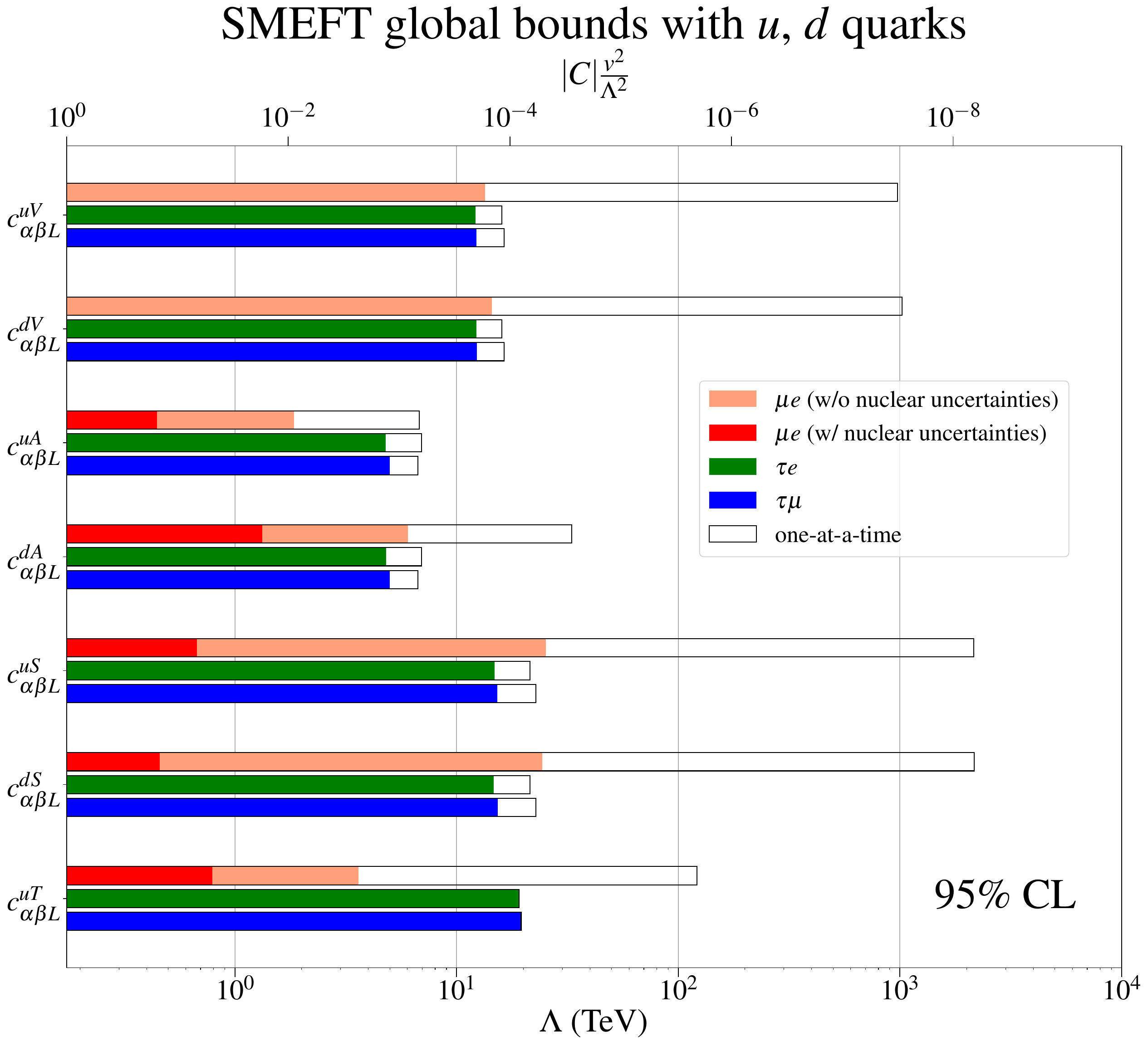}
    \caption{Current 95\% CL global bounds on the cLFV LEFT operators with only first generation quarks induced by $d=6$ SMEFT at low energies . Color code as in Fig.~\ref{fig:SMEFT_global_uds_bounds}.
    Bounds on pseudoscalar operators are equal to the scalar ones due to the correlations in Eqs.~\eqref{eq:SMEFTcorrelationsU}-\eqref{eq:SMEFTcorrelationsS}. These global bounds, along with their correlations, are collected in appendix~\ref{App:results}.}
    \label{fig:SMEFT_global_ud_bounds}
\end{figure}

As discussed in section~\ref{sec:SMEFT-tau}, when all the constraints from the different possible cLFV $\tau$ decays are considered, only a single flat direction remains unconstrained in the low energy $d=6$ SMEFT. This flat direction corresponds to a combination of the isoscalar axial and isovector scalar operators as well as the axial operator with the $s$ quark, for which only two independent constraints are available from $\tau \to \ell \eta$ and $\tau \to \ell \eta'$. If we assume operators involving only first generation quarks, these two processes are now enough to independently constrain the isoscalar axial and isovector scalar operators and no flat directions remain. Thus, in this restricted scenario, present data are enough to unambiguously constrain all cLFV operators from $d=6$ SMEFT involving first generation quarks and we display the resulting bounds from our global fit in Figure~\ref{fig:SMEFT_global_ud_bounds}.

\subsection[$\mu-e$ sector]{$\boldsymbol{\mu-e}$ sector}
\label{sec:SMEFT-mue}

In the low energy, $d=6$ SMEFT with only first generation quarks, there are $7$ relevant four-fermion operators. Thus, the dominant $8$ different constraints from $\mu-e$ conversion are in principle sufficient to simultaneously constrain all of the WCs at hand, obtaining the bounds shown in Fig.~\ref{fig:SMEFT_global_ud_bounds} with light red color.

This is no longer the case when nuclear uncertainties are properly accounted for. As previously discussed, uncertainties effectively reduce the overall number of constrained directions by $\mu-e$ conversion data. As such, $\mu-e$ conversion data will need to be supplemented by bounds coming from meson decays in order to derive constraints on all operators also in this simplified scenario.

The results of our fit taking into account nuclear uncertainties, are shown in Fig.~\ref{fig:SMEFT_global_ud_bounds} as dark red bars. The treatment of the nuclear uncertainties as nuisance parameters is equivalent to what is described in section~\ref{sec:SMEFT-s-mue}. We again find that bounds on vector and scalar operators are the most affected, as they can no longer be solely constrained through SI $\mu-e$ conversion. However, one important difference is that now the constraints have suffered from a milder degradation with respect to the case of Fig.~\ref{fig:SMEFT_global_uds_bounds}. This is a consequence of the fact that, in this simplified scenario, $\mu-e$ conversion can place constraints on $8-2=6$ combinations after nuclear uncertainties are taken into account. Therefore, in order to completely constrain the $7$ WCs, only one additional constraint coming from meson decays is necessary, contrary to the scenario with $s$-quarks, in which two extra meson constraints were necessary. As in the previous section, these constraints are very strongly correlated and are significantly different between the 1 and 2 $\sigma$ regions, when the nuisance parameters are able to reduce the number of constrained directions, as show in Fig.~\ref{fig:corr_mue_ud} in appendix~\ref{App:results}.

\section{Summary and Conclusions}
\label{sec:concl}

Charged Lepton Flavour Violation is one of our best windows to probe for generic new physics beyond the Standard Model, since the GIM suppression through the tiny neutrino masses makes these searches virtually background-free. The EFT formalism is particularly suitable to study these processes in a model-independent way. Given the large number of free parameters introduced by the EFT formalism, the most common approaches to analize their constraints are either to consider only one operator at a time or to stay at linear order in the WCs so as to explore possible flat directions that may relax the former constraints. However, the contributions of the new cLFV operators to the observables will necessarily be quadratic, as there are no SM contributions to interfere with, and therefore going beyond the one-operator-at-a-time approach in these studies is challenging.

In this work we have focused on the impact of potential flat directions and on studying how the bounds obtained from a one-operator-at-a-time may be affected by them. After introducing the LEFT formalism and its matching to the $d=6$ SMEFT, we analized the bounds from cLFV lepton and meson decays as well as from $\mu - e$ conversion in nuclei on the WCs directly at the low-energy scale relevant for the most important observables.

We find that, for the dipole and fully leptonic operators, there are no flat directions that may relax the bounds that can be derived simply through the one-operator-at-a-time approach from processes such as $\ell_\alpha \to \ell_\beta \gamma$ and $\ell_\alpha \to \ell_\beta \ell_\beta \ell_\gamma$, respectively. While there are some operators that are not bounded by these processes and that we list, these do not hinder the constraints derived on the others which are summarized in Fig.~\ref{fig:circular_barplot}. 

Conversely, for semileptonic 4-fermion operators the situation is very different. It is also very different for cLFV involving the $\tau$ with respect to the $\mu-e$ case. The former is mainly bounded by searches of $\tau$ decays to a lighter lepton and a meson. The different isospin and nature of the meson gives sensitivity to operators with different quark content and Lorentz structure. Nevertheless, and despite the many different and complementary searches, we find that in the most general case there are several flat directions that relax some of the constraints that would be obtained through the one-operator-at-a-time approach. In particular, our LEFT scenario has 15 coefficients, 5 per Lorentz structure (tensor, vector, axial, scalar and pseudoscalar) and 3 per quark type (up, down and strange), that may contribute. While through a one-operator-at-a-time approach all WCs can be bounded at the level of $10^{-3}-10^{-4}$ (see Fig.~\ref{fig:one-at-a-time_bounds}), we find that these constraints apply in a fully global analysis for only 7 of the 15 operators (see Fig.~\ref{fig:LEFT_global_bounds}), corresponding to the vector and tensor structures as well as the scalar one for the $s$ quarks. Among the remaining 8 operators, 4 flat directions exist so that individual bounds on their WCs cannot be derived in a global analysis as one may freely move along the unconstrained flat directions.

Given this situation, we then analize the more constrained scenario of the $d=6$ SMEFT operators at low energy. This situation is described with only 10 instead of the 15 parameters of the LEFT scenario. Nevertheless, as shown in Fig.~\ref{fig:SMEFT_global_uds_bounds}, the only global bounds are still those derived for the vector and tensor structures as well as for the scalar operator with the $s$ quark. Indeed, there is a single flat direction involving the other 5 operators remaining. Thus, even though 9 parameter combinations out of the 10 independent WCs are bounded down to $\sim 10^{-4}$, Fig.~\ref{fig:SMEFT_global_uds_bounds} does not display a bound for 5 of them since they are all involved in the unconstrained direction. Adding $\tau\to\ell KK$ would close this remaining flat direction and lead to global constrains for all WCs, nevertheless a better handle on its form factors is needed in order to properly include it in the analysis.
The final simplified scenario we consider is when only first generation quarks participate in the observables. In this case, as shown in Fig.~\ref{fig:SMEFT_global_ud_bounds}, all WCs are constrained also in the global analysis with bounds ranging from $10^{-3}$ to $10^{-4}$.   

The $\mu-e$ sector for the leptonic operators is the most complex to analyze, since the would-be flat directions are determined by overlap integrals and nuclear parameters defining the particular combination of operators that contribute to the SI and SD transitions of each nuclei. Moreover, uncertainties on these quantities, when incorporated as nuisance parameters to a global fit, alter the constrained directions to the point that some become linearly dependent on others. Indeed, within uncertainties, the directions probed by Pb and S become parallel to those determined by Au and Ti respectively, reducing the 4 independent parameter combinations probed by the different elements to only 2. As such, additional data on more complementary nuclei would be very helpful~\cite{Davidson:2018kud}. We notice, however, that SI transitions are fully characterized by only 4 operator combinations: the scalar and vector couplings to protons and neutrons. Thus, once 4 independent operator combinations corresponding to 4 different nuclei have been bounded, new nuclei cannot provide complementary information. The situation is similar for the SD contribution to $\mu-e$ conversion. Again only 4 operator combinations (isoscalar and isovetor transverse and longitudinal modes) contribute. Furthermore, present data on $\mu-e$ conversion in Ti already provide independent constraints on all 4, so that additional data will not allow to constrain new directions. Conversely, we find that cLFV meson decays such as $\pi^0 \to \mu e$ or $\eta \to \mu e$ do provide complementary information, although the bounds are much weaker. Thus, improving these constraints does have a significant impact in global fits when correlations and flat directions are fully accounted for.

With the one-operator-at-a-time approach and neglecting nuisance parameters, very stringent bounds down to $10^{-7}$ are found for the LEFT vector and scalar operators that contribute directly to the SI $\mu-e$ conversion in nuclei. Tensor structures have somewhat weaker bounds of $10^{-5}$ through their finite recoil contribution to the SI transition. The axial and pseudoscalar structures contribute to SD transitions instead, which lack the resonant enhancement of the SI and lead to bounds between $10^{-3}$ and $10^{-5}$ (see Fig.~\ref{fig:one-at-a-time_bounds}).  None of the observables are sensitive to the vector operator involving the $s$ quark, which remains unbounded. When analizing the LEFT scenario with its 15 free parameters, only bounds on the vector and pseudoscalar operators may be derived. For the vectors they are relaxed from $10^{-7}$ to $10^{-3}$ due to the very many flat directions present (see Fig.~\ref{fig:LEFT_global_bounds}). The bounds on the pseudoscalar operators are now dominated by the $\eta \to \mu e$ process and relaxed by around 5 orders of magnitude.

When the low energy $d=6$ SMEFT with its 10 operators is considered instead, all of these flat directions are lifted. However, there is a single direction involving the axial operators in an isoscalar combination as well as the $s$ quark contribution which is only very weakly constrained, beyond the range of Fig.~\ref{fig:SMEFT_global_uds_bounds}, by $\eta \to \mu e$. The bounds on the other operators are also very degraded with respect to the one-at-a-time constraints since, even when independent, many of the directions constrained by $\mu-e$ conversion are almost parallel. Thus, from Fig.~\ref{fig:one-at-a-time_bounds} to Fig.~\ref{fig:SMEFT_global_uds_bounds} the constraints weaken by $\sim 4$ orders of magnitude in the global SMEFT analysis. The situation worsens when nuclear uncertainties are accounted for and SI $\mu-e$ conversion can effectively constrain only 2 directions. Therefore, the bounds on the vector and scalar operators become only order 1 or even larger for the former.

Finally, with the additional simplifying assumption of no operators involving $s$ quarks, bounds on all WCs may be derived through the global analysis. Without nuclear uncertainties, these bounds on the $\mu-e$ sector are surprisingly similar to those on the $\tau-\ell$ sector. While with the one-at-a-time analysis much stronger constraints in the $\mu-e$ were found, in a global fit scenario, they become diluted by the very strong correlations present between the different observables. Moreover, when nuclear uncertainties are included as nuisance parameters, the bounds on the $\mu-e$ sector become significantly weaker than those in the $\tau-\ell$, particularly the ones for the vector and scalar operators, as previously discussed.

All in all, we find that flat directions play no role in the bounds on fully leptonic operators and the naive one-operator-at-a-time approach leads to reliable constraints on the relevant WCs. Conversely, flat directions appear and lead to fully unconstrained parameter combinations for semileptonic cLFV 4-fermion operators. While these flat directions were not found in previous global scans of the parameter space, we believe this is due to our different scanning strategy as outlined in section~\ref{sec:LEFT}. Remarkably, we find that when the operators are those induced by the low-energy $d=6$ SMEFT and if only two independent couplings (one for up and one for down quarks) are considered for each Lorentz structure, present data allows to lift all flat directions present and obtain unambiguous bounds on all operators. However, in the case of the $\mu-e$ sector, these bounds are around 4 magnitude weaker than in the one-operator-at-a-time approach and become even weaker when nuclear uncertainties are properly accounted for. Nevertheless, extremely strong correlations among them exist, reflecting the underlying directions that are much more stringently constraint. Thus, we also provide correlation matrices in a~\href{https://github.com/dnaredo/cLFV_GlobalBounds.git}{GitHub repository} that contain all these nuances as a useful tool to incorporate cLFV constraints in particular UV-complete scenarios.

\medskip
\paragraph{Acknowledgments.}
The authors thank Javier Menéndez, Jacobo López-Pavón and Salvador Urrea for very illuminating discussions, and Xavier Ponce Díaz for pointing out a typo in our numerical analysis of the three-body leptonic decays.
This project has received support from the European Union’s Horizon 2020 research and innovation programme under the Marie Skłodowska-Curie grant agreement No~860881-HIDDeN and No 101086085 - ASYMMETRY, and from the Spanish Research Agency (Agencia Estatal de Investigaci\'on) through the Grant IFT Centro de Excelencia Severo Ochoa No CEX2020-001007-S and Grant PID2022
137127NB-I00 funded by MCIN/AEI/10.13039/501100011033. 
We acknowledge support from the HPC-Hydra cluster at IFT.
XM acknowledges funding from the European Union’s Horizon Europe Programme under the Marie Skłodowska-Curie grant agreement no.~101066105-PheNUmenal. The work of DNT was supported by the
Spanish MIU through the National Program FPU (grant number FPU20/05333).

\appendix

\section{Detailed fit results}
\label{App:results}

Here we present quantitative information about our bounds and the correlations among the different parameters so as to allow the implementation of our constraints in specific scenarios.

Correlations are of particular importance for the global fit results, specially for the $\mu-e$ global bounds, which get substantially degraded with respect to the one-at-a-time scenario. This is due to the fact that some specific combinations of WCs are very poorly constrained (or not at all) together with extremely tight bounds in other very specific directions. However, if a particular UV completion does not align along these weakly constrained directions, the corresponding constraints will be much tighter than those directly inferred from the plots showed in the previous sections. 

Usually, this information is easily conveyed through the covariance matrix. However, all the observables considered, being cLFV and therefore not present in the SM, depend quadratically on the WCs at the leading order. Furthermore, as there is no signal in any of the observables, the best-fit point corresponds to all WCs vanishing and the resulting $\chi^2$ test-statistics will be a purely quartic polynomial.

Indeed, if one tries to approximate the test-statistics from its Taylor's series around the best-fit point, the first non-vanishing order will be the quartic:
\begin{align}
    \chi^2\left(c_i\right)&\simeq\chi^2\left(c_i=0\right)+\dfrac{c_i c_j}{2!}\dfrac{\partial^2 \chi^2}{\partial c_i\partial c_j}\Bigg\rvert_{c_i=0}+\dfrac{c_i c_k c_k}{3!}\dfrac{\partial^3 \chi^2}{\partial c_i\partial c_j \partial c_k}\Bigg\rvert_{c_i=0}+\dfrac{c_i c_j c_k c_l}{4!}\dfrac{\partial^4 \chi^2}{\partial c_i\partial c_j \partial c_k\partial c_l}\Bigg\rvert_{c_i=0}\nonumber\\
    &=\dfrac{c_i c_j c_k c_l}{4!}\dfrac{\partial^4 \chi^2}{\partial c_i\partial c_j \partial c_k\partial c_l}\Bigg\rvert_{c_i=0}\,,
\end{align}
and thus the covariance matrix vanishes for our test-statistics, such that the information about possible correlations between the coefficients will actually be contained within the co-kurtosis tensor of 4th derivatives.

Ideally, a change of basis to the second order polynomials in the WC relevant for the observables considered would allow to make the $\chi^2$ a quadratic function of the parameters analized. Unfortunately, after doing this change of basis for the most relevant observables, it is not possible to express the remaining WC combinations necessary for the rest of the measurements as a function of the new variables in an univocal way. Thus, for the sake of presenting results that can be more easily implemented when deriving constraints on specific models and that approximate the results of our global fit, we construct a ``proxy-covariance matrix'' whose global bounds match those extracted from the full analysis of the true test-statistics at \emph{a given confidence level}.

In particular, we construct a ``proxy-test-statistics'' by using as observables the square root of the branching ratio, instead of the branching ratio itself. This guarantees that our ``proxy-test-statistics'' is a quadratic polynomial in the coefficients and consequently has a well-defined covariance matrix. We then normalise the covariance matrix so as to only retain information on how correlated are the WCs. Lastly, the normalisation of the variances of each of the WCs is set to the global bound obtained in the full analysis of the true test-statistics at some given confidence level. Thus, the information about the strongly correlated directions is preserved and the bounds agree with those of the actual global fit at the confidence level selected. Notice, however, that using this covariance matrix for different confidence levels would provide wrong results as the dependence on the WCs of the true test statistics (quartic) and the approximated one (quadratic) is different. 

In particular, we will give our results in terms of the global bounds $\{\sigma_i\}$ on the coefficients $\{c_i\}$, with a normalized ``proxy-covariance matrix'' $\rho$, such that the ``proxy-test-statistics'' can be easily constructed as follows:
\begin{equation}
    \chi^2_{\text{proxy}}\equiv \sum_{i,j}\left(\rho^{-1}\right)_{ij}\frac{c_i}{\sigma_i}\frac{c_j}{\sigma_j}\,.
\end{equation}

The bounds $\{\sigma_i\}$ and their corresponding inverse-covariances $\rho^{-1}$ are available in text file format in the following~\href{https://github.com/dnaredo/cLFV_GlobalBounds.git}{GitHub repository}.

\subsection[$\tau-\ell$ SMEFT results]{$\boldsymbol{\tau-\ell}$ SMEFT results}
\subsubsection{Results with $u$, $d$ and $s$ quarks}
\begin{figure}[t!]
    \centering
    \includegraphics[width=\textwidth]{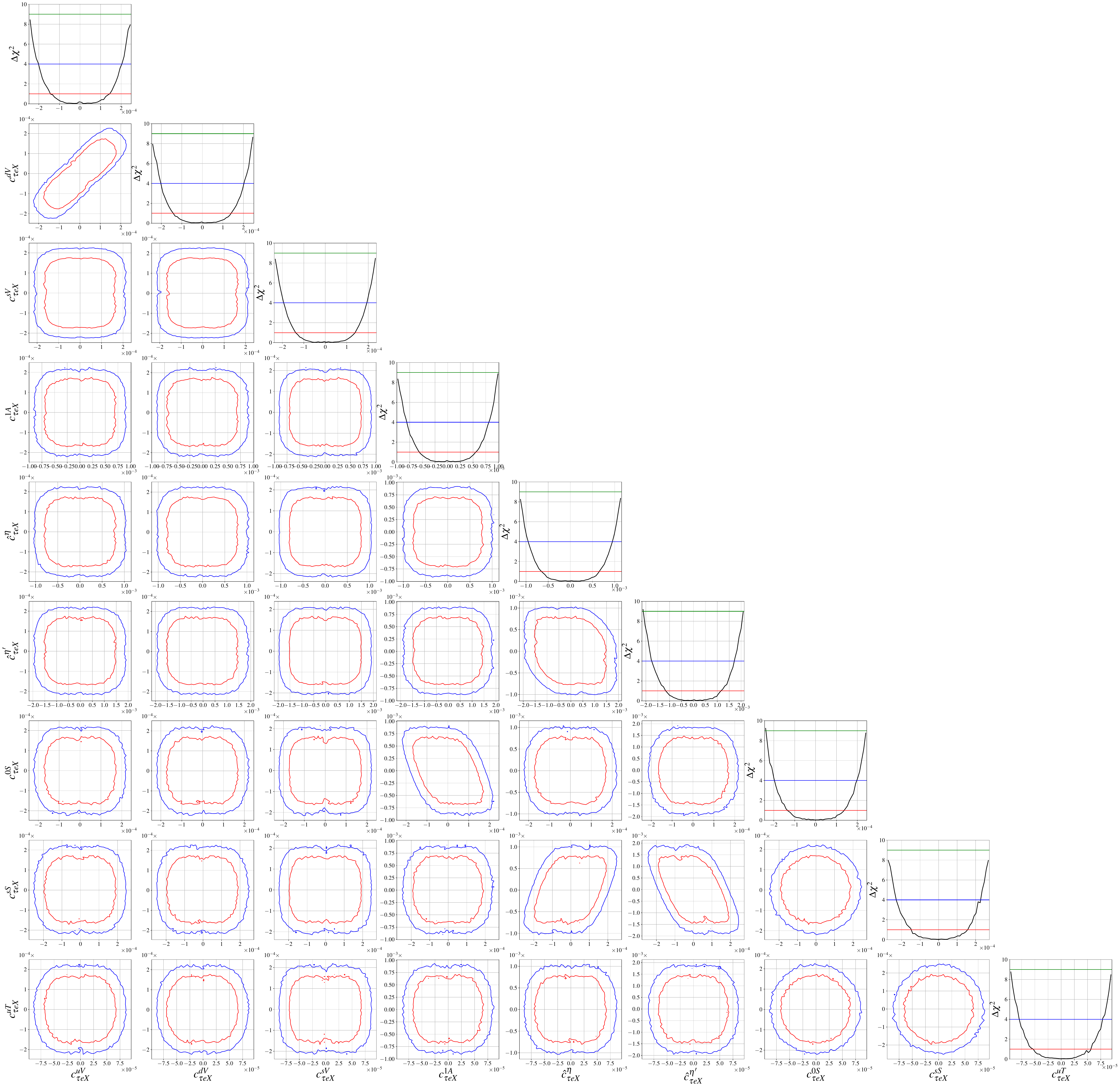}
    \caption{Correlations between the different $\tau-e$ operators, as extracted from our numerical analysis of the LEFT from $d=6$ SMEFT at low energy, considering all light quarks. Red and blue lines correspond to 68\% and 95\% C.L., respectively. Notice the generally mild correlations between the coefficients, which explain why the global bounds (except in the presence of flat directions) are so close to the one-at-a-time bounds (see Figs.~\ref{fig:SMEFT_global_uds_bounds} and~\ref{fig:SMEFT_global_ud_bounds}). A very similar plot is obtained for the $\tau-\mu$ sector.}
    \label{fig:corr_taue}
\end{figure}

As argued in section~\ref{sec:SMEFT}, when considering all light quarks, there is a single flat direction remaining, which we isolate by defining the following uncorrelated coefficient combinations:
\begin{align}
    c^{1A}_{\tau\ell X}&=c^{uA}_{\tau\ell X}-c^{dA}_{\tau\ell X}\,,\\
    c^{0S}_{\tau\ell X}&=c^{uS}_{\tau\ell X}+c^{dS}_{\tau\ell X}\,,\\
    \hat{c}_{\tau\ell X}^{\eta}&=\dfrac{f_{\eta}^{u}}{f_\pi}\left(c^{uA}_{\tau\ell X}+c^{dA}_{\tau\ell X}\right)+\dfrac{f_{\eta}^{s}}{f_\pi}c^{sA}_{\tau\ell X}\mp \dfrac{h^{u}_\eta}{f_\pi m_\tau (m_u+m_d)}\left(c^{uS}_{\tau\ell X}-c^{dS}_{\tau\ell X}\right)\,,\\
    \hat{c}_{\tau\ell X}^{\eta'}&=\dfrac{f_{\eta'}^{u}}{f_\pi}\left(c^{uA}_{\tau\ell X}+c^{dA}_{\tau\ell X}\right)+\dfrac{f_{\eta'}^{s}}{f_\pi}c^{sA}_{\tau\ell X}\mp \dfrac{h^{u}_{\eta'}}{f_\pi m_\tau (m_u+m_d)}\left(c^{uS}_{\tau\ell X}-c^{dS}_{\tau\ell X}\right)\,,
\end{align}
where the $\mp$ sign corresponds to $X=L/R$, respectively. 
At $95\%$ CL, the global bounds read,
\begin{equation}
    \begin{pmatrix}
        c^{uV}_{\tau eX}\\\\
        c^{dV}_{\tau eX}\\\\
        c^{sV}_{\tau eX}\\\\
        c^{1A}_{\tau eX}\\\\
        \hat{c}^{\eta}_{\tau e X}\\\\
        \hat{c}^{\eta'}_{\tau e X}\\\\
        c^{0S}_{\tau eX}\\\\
        c^{sS}_{\tau eX}\\\\
        c^{uT}_{\tau eX}
    \end{pmatrix}
    <
    \begin{pmatrix}
        2.0\times 10^{-4}\\\\
        2.0\times 10^{-4}\\\\
        1.9\times 10^{-4}\\\\
        7.9\times 10^{-4}\\\\
        9.3\times 10^{-4}\\\\
        1.7\times 10^{-3}\\\\
        1.9\times 10^{-4}\\\\
        2.3\times 10^{-4}\\\\
        7.7\times 10^{-5}
    \end{pmatrix}
    \,,\hspace{2cm}
    \begin{pmatrix}
        c^{uV}_{\tau \mu X}\\\\
        c^{dV}_{\tau \mu X}\\\\
        c^{sV}_{\tau \mu X}\\\\
        c^{1A}_{\tau \mu X}\\\\
        \hat{c}^{\eta}_{\tau \mu X}\\\\
        \hat{c}^{\eta'}_{\tau \mu X}\\\\
        c^{0S}_{\tau \mu X}\\\\
        c^{sS}_{\tau \mu X}\\\\
        c^{uT}_{\tau \mu X}
    \end{pmatrix}
    <
    \begin{pmatrix}
        2.0\times 10^{-4}\\\\
        2.0\times 10^{-4}\\\\
        3.0\times 10^{-4}\\\\
        8.9\times 10^{-4}\\\\
        8.0\times 10^{-4}\\\\
        1.5\times 10^{-3}\\\\
        1.8\times 10^{-4}\\\\
        2.2\times 10^{-4}\\\\
        7.8\times 10^{-5}
    \end{pmatrix}
    \,,
\end{equation}
which are mostly uncorrelated, as shown in Fig.~\ref{fig:corr_taue}.

\subsubsection{Results with first generation quarks}
At $95\%$ CL, the global bounds read,
\begin{equation}
    \begin{pmatrix}
        c^{uV}_{\tau eX}\\\\
        c^{dV}_{\tau eX}\\\\
        c^{uA}_{\tau eX}\\\\
        c^{dA}_{\tau eX}\\\\
        c^{uS}_{\tau eX}\\\\
        c^{dS}_{\tau eX}\\\\
        c^{uT}_{\tau eX}
    \end{pmatrix}
    <
    \begin{pmatrix}
        2.0\times 10^{-4}\\\\
        2.0\times 10^{-4}\\\\
        1.3\times 10^{-3}\\\\
        1.3\times 10^{-3}\\\\
        1.3\times 10^{-4}\\\\
        1.4\times 10^{-4}\\\\
        8.1\times 10^{-5}
    \end{pmatrix}
    \,,\hspace{2cm}
    \begin{pmatrix}
        c^{uV}_{\tau \mu X}\\\\
        c^{dV}_{\tau \mu X}\\\\
        c^{uA}_{\tau \mu X}\\\\
        c^{dA}_{\tau \mu X}\\\\
        c^{uS}_{\tau \mu X}\\\\
        c^{dS}_{\tau \mu X}\\\\
        c^{uT}_{\tau \mu X}
    \end{pmatrix}
    <
    \begin{pmatrix}
        2.0\times 10^{-4}\\\\
        2.0\times 10^{-4}\\\\
        1.2\times 10^{-3}\\\\
        1.2\times 10^{-3}\\\\
        1.3\times 10^{-4}\\\\
        1.3\times 10^{-4}\\\\
        7.6\times 10^{-5}
    \end{pmatrix}
    \,.
\end{equation}

\vspace{1cm}
\subsection[$\mu-e$ SMEFT results]{$\boldsymbol{\mu-e}$ SMEFT results}
As argued in sections~\ref{sec:SMEFT} and~\ref{sec:SMEFT-s}, all WCs under consideration, except for $c^{sV}_{\mu e X}$, can be simultaneously constrained in our global fit. We collect these global bounds here for the analyses including the nuclear uncertainties, as well as their correlations showed in Fig.~\ref{fig:corr_mue_uds} and Fig.~\ref{fig:corr_mue_ud}.

We note that, contrary to the $\tau$ sector, the correlation between some of the operators is extremely strong. This fact explains why the global bounds are so much weaker compared with the the one-at-a-time scenario (see Fig.~\ref{fig:SMEFT_global_uds_bounds}), pushing some of the profiled bounds to $\mathcal O(\gg1)$.
Nevertheless, these strong correlations imply that these very weak bounds can only be saturated in models that predict very specific cancellations.
Saturating all of them would in general result in incorrect results, underestimating the cLFV bounds, and thus we instead provide the correlation matrix $\rho^{-1}$ available in the~\href{https://github.com/dnaredo/cLFV_GlobalBounds.git}{GitHub repository} in order to correctly include these bounds on specific setups.

We also remark the big difference between the 68\% C.L and 95\% C.L. bounds, being the former orders of magnitude stronger for some operators. The reason, as explained before, is that at higher C.L. two constraints from SI $\mu-e$ conversion are lost due to nuclear uncertainties in the overlap integrals defining the constrained directions (see Eq.~\eqref{eq:mueconvSI}).
Therefore, the 95\% C.L. global bounds presented in this work cannot be naively translated into other confidence levels and the complete $\chi^2$ must be used. 

In the following we present bounds and correlations extracted for the operators in which the electron has $L$-chirality, but very similar results are found in the $R$-chiral case, which are available in the repository.

\newpage
\subsubsection{Results with $u$, $d$ and $s$ quarks and nuclear uncertainties}
\begin{figure}[h!]
    \centering
    \includegraphics[width=\textwidth]{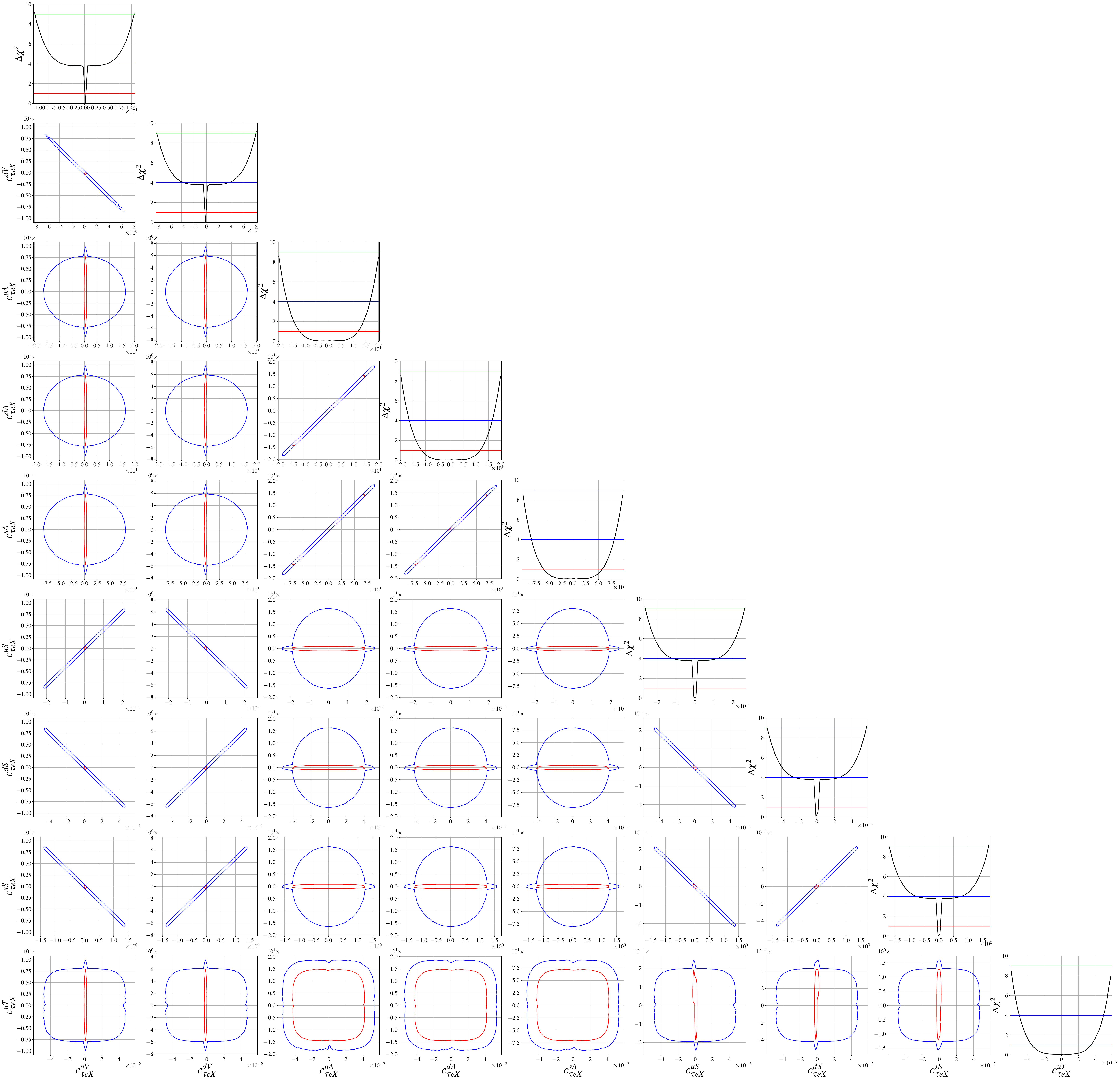}
    \caption{Same as Fig.~\ref{fig:corr_taue} but for the $\mu-e$ sector, showing the very strong correlations between some of the coefficients. This explains the huge relaxation of the global bounds with respect to the one-at-a-time scenario in Fig.~\ref{fig:SMEFT_global_uds_bounds}.}
    \label{fig:corr_mue_uds}
\end{figure}

At $95\%$ CL, the global bounds read,

\begin{equation}
    \begin{pmatrix}
        c^{uV}_{\mu e L}\\\\
        c^{dV}_{\mu e L}\\\\
        c^{uA}_{\mu e L}\\\\
        c^{dA}_{\mu e L}\\\\
        c^{sA}_{\mu e L}\\\\
        c^{uS}_{\mu e L}\\\\
        c^{dS}_{\mu e L}\\\\
        c^{sS}_{\mu e L}\\\\
        c^{uT}_{\mu e L}
    \end{pmatrix}
    <
    \begin{pmatrix}
        4.6\times10^{0}\\\\
        3.5\times10^{0}\\\\
        1.6\times 10^{1}\\\\
        1.6\times10^{1}\\\\
        7.8\times10^{1}\\\\
        1.1\times10^{-1}\\\\
        2.4\times10^{-1}\\\\
        7.4\times10^{-1}\\\\
        4.8\times10^{-2}
    \end{pmatrix},
\end{equation}
with the strong correlations shown in Fig.~\ref{fig:corr_mue_uds}.

\subsubsection{Results with first generation quarks and nuclear uncertainties}
\begin{figure}[h!]
    \centering
    \includegraphics[width=\textwidth]{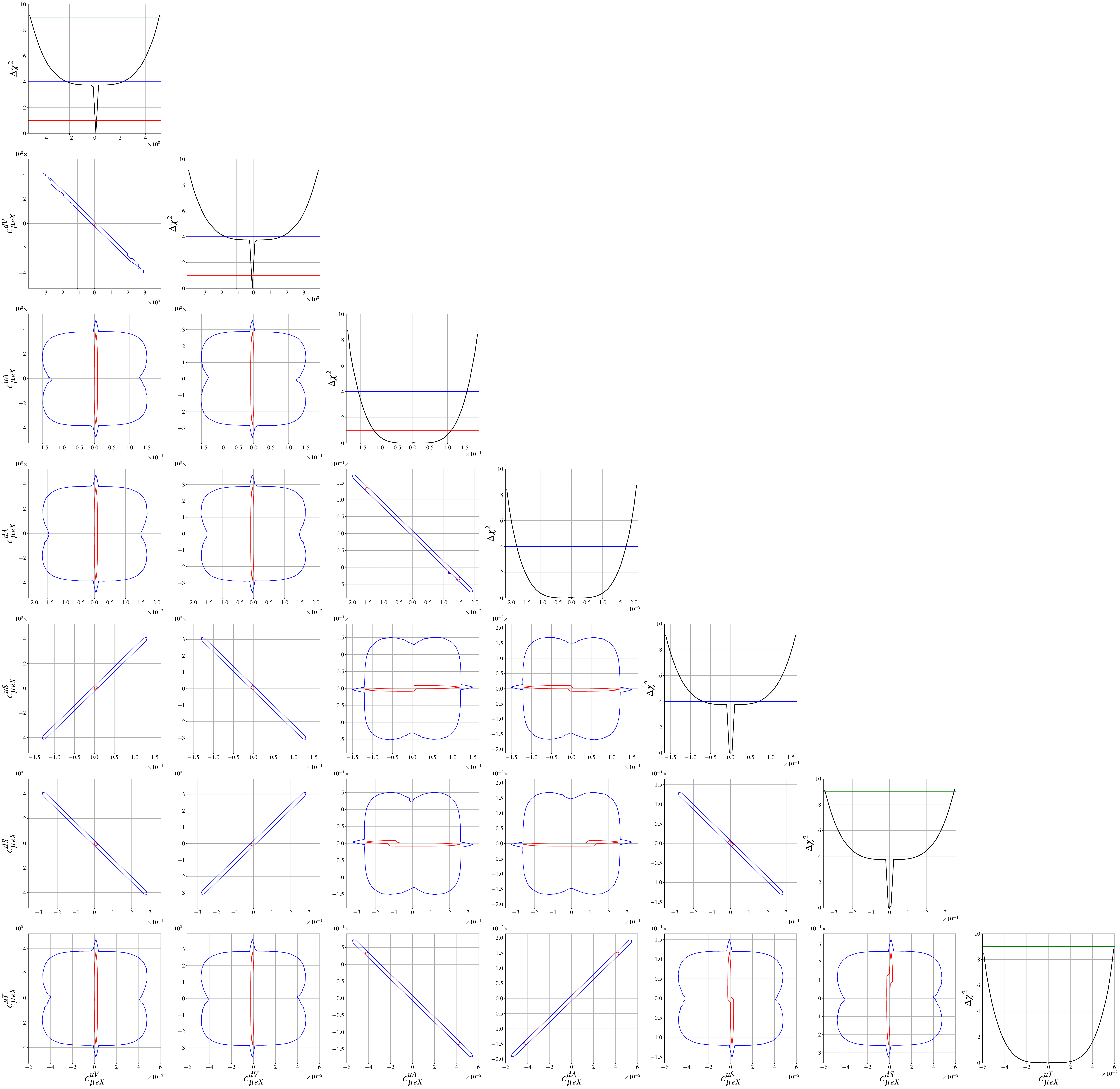}
    \caption{Same as Fig.~\ref{fig:corr_mue_uds}, but for the scenario with only first generation quarks. This explains the huge relaxation of the global bounds with respect to the one-at-a-time scenario in Fig.~\ref{fig:SMEFT_global_ud_bounds}.}
    \label{fig:corr_mue_ud}
\end{figure}

Global bounds at 95\% CL:

\begin{equation}
    \begin{pmatrix}
        c^{uV}_{\mu e L}\\\\
        c^{dV}_{\mu e L}\\\\
        c^{uA}_{\mu e L}\\\\
        c^{dA}_{\mu e L}\\\\
        c^{uS}_{\mu e L}\\\\
        c^{dS}_{\mu e L}\\\\
        c^{uT}_{\mu e L}
    \end{pmatrix}
    <
    \begin{pmatrix}
        2.1\times10^{0}\\\\
        1.6\times10^{0}\\\\
        1.5\times10^{-1}\\\\
        1.7\times10^{-2}\\\\
        6.7\times10^{-2}\\\\
        1.4\times10^{-1}\\\\
        4.9\times10^{-2}
    \end{pmatrix},
\end{equation}
with the strong correlations shown in Fig.~\ref{fig:corr_mue_ud}.

\newpage

\newpage

\newpage

\section{Bounds on dipole and four-lepton operators}
\label{app:dipole_4l_bounds}
For the sake of completeness, we also show the bounds on the LEFT operators as extracted from radiative and three-body leptonic decays. As discussed in section~\ref{sec:lepobs}, the incoherent contributions of each of the operators renders the following 95\% C.L. bounds totally uncorrelated:
\begin{equation}
\begin{pmatrix}
    c_{e\mu}^{e\gamma}\\\\
    c_{e\tau}^{e\gamma}\\\\
    c_{\mu\tau}^{e\gamma}
\end{pmatrix}
<
\begin{pmatrix}
  6.0\times 10^{-12}\\\\
  7.9\times 10^{-8}\\\\
  8.9\times 10^{-8}
\end{pmatrix}\,,
\hspace{1.5cm}
\begin{pmatrix}
    c_{e\mu L}^{eeLV}\\\\
    c_{e\tau L}^{eeLV}\\\\
    c_{\mu\tau L}^{\mu\mu LV}\\\\
    c_{e\tau L}^{\mu\mu LV}\\\\
    c_{\mu\tau L}^{eeLV}\\\\
    c_{e\tau L}^{e\mu LV}\\\\
    c_{\mu\tau L}^{\mu eLV}
\end{pmatrix}
<
\begin{pmatrix}
  7.8\times 10^{-7}\\\\
  3.0\times 10^{-4}\\\\
  2.7\times 10^{-4}\\\\
  4.3\times 10^{-4}\\\\
  4.3\times 10^{-4}\\\\
  2.3\times 10^{-4}\\\\
  2.4\times 10^{-4}
\end{pmatrix}
\,,
\end{equation}

\begin{equation}
\begin{pmatrix}
    c_{e\mu L}^{eeRV}\\\\
    c_{e\tau L}^{eeRV}\\\\
    c_{\mu\tau L}^{\mu\mu RV}\\\\
    c_{e\tau L}^{\mu\mu RV}\\\\
    c_{\mu\tau L}^{e\mu RV}\\\\
    c_{\mu\tau L}^{eeRV}\\\\
    c_{e\tau L}^{\mu eRV}\\\\
    c_{e\tau L}^{e\mu RV}\\\\      
    c_{\mu\tau L}^{\mu eRV}
\end{pmatrix}
<
\begin{pmatrix}
  1.1\times 10^{-6}\\\\
  4.3\times 10^{-4}\\\\
  3.8\times 10^{-4}\\\\
  4.3\times 10^{-4}\\\\
  4.3\times 10^{-4}\\\\
  4.3\times 10^{-4}\\\\
  4.3\times 10^{-4}\\\\
  3.2\times 10^{-4}\\\\
  3.4\times 10^{-4}
\end{pmatrix}\,,
\hspace{1.5cm}
\begin{pmatrix}
    c_{e\mu R}^{eeRS}\\\\
    c_{e\tau R}^{eeRS}\\\\
    c_{\mu\tau R}^{\mu\mu RS}\\\\
    c_{e\tau R}^{\mu\mu RS}\\\\
    c_{\mu\tau R}^{e\mu RS}\\\\
    c_{\mu\tau R}^{eeRS}\\\\
    c_{e\tau R}^{\mu eRS}\\\\
    c_{e\tau R}^{e\mu RS}\\\\
    c_{\mu\tau R}^{\mu eRS}
\end{pmatrix}
<
\begin{pmatrix}
  3.1\times 10^{-6}\\\\
  1.2\times 10^{-3}\\\\
  1.1\times 10^{-3}\\\\
  8.6\times 10^{-4}\\\\
  8.6\times 10^{-4}\\\\
  8.6\times 10^{-4}\\\\
  8.6\times 10^{-4}\\\\
  9.0\times 10^{-4}\\\\
  9.6\times 10^{-4}
\end{pmatrix}\,.
\end{equation}
The same exact bounds apply to the WCs obtained from interchanging $(L\longleftrightarrow R)$.


\bibliographystyle{JHEP} 
\bibliography{biblio}

\end{document}